\documentclass[a4paper,11pt]{article}

\pdfoutput=1

\usepackage{amsmath,amssymb,mathtools}
\usepackage{color}
\usepackage{graphicx}
\usepackage{subfigure}
\usepackage{cite}
\usepackage[colorlinks=true,linkcolor=blue,citecolor=red,urlcolor=blue,bookmarks]{hyperref}
\usepackage{multirow,makecell}
\usepackage{textcomp}
\usepackage{wasysym}
\usepackage{ulem}

\usepackage[utf8]{inputenc}
\usepackage[T1]{fontenc}

\usepackage[text={17.1cm,24.6cm},centering]{geometry} 

\newcommand{\mr}[2]{\multirow{#1}*{#2}}
\newcommand{\mc}[3]{\multicolumn{#1}{#2}{#3}}

\numberwithin{equation}{section}

\def \be {\begin{equation}}
\def \ee {\end{equation}}
\def \ba {\begin{array}}
\def \ea {\end{array}}
\def \bea{\begin{eqnarray}}
\def \eea{\end{eqnarray}}
\def \nn {\nonumber}

\def \a {\alpha}
\def \b {\beta}
\def \g {\gamma}

\def \G {\Gamma}
\def \d {\delta}
\def \D {\Delta}
\def \e {\epsilon}
\def \ve {\varepsilon}

\def \n {\nu}

\def \lam {\lambda}
\def \Lam {\Lambda}
\def \s {\sigma}

\def \r {\rho}

\def \O {\Omega}
\def \th {\theta}

\def \io {\iota}

\def \bs {\backslash}

\def \bA {\mathbf A}
\def \bB {\mathbf B}

\def \cA {\mathcal A}
\def \cB {\mathcal B}

\def \cI {\mathcal I}
\def \cJ {\mathcal J}
\def \cK {\mathcal K}

\def \cN {\mathcal N}

\def \cP {\mathcal P}
\def \cQ {\mathcal Q}
\def \cR {\mathcal R}
\def \cS {\mathcal S}

\def \cU {\mathcal U}
\def \cV {\mathcal V}
\def \cW {\mathcal W}
\def \cX {\mathcal X}
\def \cY {\mathcal Y}
\def \cZ {\mathcal Z}

\def \rZ {\mathrm Z}

\def \f {\frac}

\def \lt {\left}
\def \rt {\right}

\def \sr {\sqrt}
\def \td {\tilde}

\def \inf {\infty}

\def \lag {\langle}
\def \rag {\rangle}

\def \ep {\mathrm{e}}
\def \ii {\mathrm{i}}

\def \tr {\textrm{tr}}

\def \diag {\mathop{\textrm{diag}}}

\def \and {{~\textrm{and}~}}

\def \per {\mathop{\textrm{per}}}

\def \bos {\textrm{bos}}
\def \fer {\textrm{fer}}
\def \XXX {\textrm{XXX}}
\def \XXZ {\textrm{XXZ}}

\begin{document}

\title{
\textbf{Entanglement of magnon excitations in spin chains}
}
\author{
Jiaju Zhang$^{1}$ 
and
M. A. Rajabpour$^{2}$ 
}
\date{}
\maketitle
\vspace{-10mm}
\begin{center}
{\it
$^{1}$Center for Joint Quantum Studies and Department of Physics, School of Science,\\
      Tianjin University, 135 Yaguan Road, Tianjin 300350, China\\\vspace{1mm}
$^{2}$Instituto de Fisica, Universidade Federal Fluminense,\\
      Av. Gal. Milton Tavares de Souza s/n, Gragoat\'a, 24210-346, Niter\'oi, RJ, Brazil
}
\vspace{10mm}
\end{center}

\begin{abstract}
  We calculate exactly the entanglement content of magnon excited states in the integrable spin-1/2 XXX and XXZ chains in the scaling limit. In particular, we show that as far as the number of excited magnons with respect to the size of the system is small one can decompose the entanglement content, in the scaling limit, to the sum of the entanglement of particular excited states of free fermionic or bosonic theories. In addition we conjecture that the entanglement content of the generic translational invariant free fermionic and bosonic Hamiltonians can be also classified, in the scaling limit, with respect to the entanglement content of the fermionic and bosonic chains with the number operator as the Hamiltonian in certain circumstances. Our results effectively classify the entanglement content of wide range of integrable spin chains in the scaling limit.
\end{abstract}

\baselineskip 18pt
\thispagestyle{empty}
\newpage


\tableofcontents

\section{Introduction}

Entanglement between subsystems in extended many-body systems has become one of the key ingredients to understanding of quantum matter \cite{Amico:2007ag,Eisert:2008ur,Calabrese:2009bph,Laflorencie:2015eck,Witten:2018lha}.
A quantitative characterization of quantum entanglement is the entanglement entropy and it is defined as follows:
For a quantum system in state $|K\rag$, one can divide the whole system into two subsystems $A$ and $B$, integrate out the degrees of freedom of the subsystem $B$ of the total system density matrix $\r_K=|K\rag\lag K|$, and obtain the reduced density matrix (RDM) $\r_{A,K}=\tr_B\r_K$ of the subsystem $A$.
The entanglement entropy is the von Neumann entropy of the RDM
\be
S_{A,K}=-\tr_A(\r_{A,K}\log\r_{A,K}).
\ee
It can be calculated as the $n\to1$ limit of the R\'enyi entropy
\be
S_{A,K}^{(n)}=-\f{1}{n-1}\log\tr_A\r_{A,K}^n.
\ee
The entanglement entropy and R\'enyi entropy in various extended quantum systems have been investigated for the ground state
\cite{Bombelli:1986rw,Srednicki:1993im,Callan:1994py,Holzhey:1994we,Peschel:1998ftd,Peschel:1999pkr,%
Chung:2000tqg,Chung:2001oyk,Cheong:2002ukf,Vidal:2002rm,Peschel:2002jhw,Latorre:2003kg,Jin:2004pgk,Korepin:2004zz,%
Plenio:2004he,Calabrese:2004eu,Cramer:2005mx,Casini:2005rm,Casini:2005zv,Casini:2009sr,Calabrese:2009qy,Peschel:2009iuj,%
Peschel:2012jed},
 the low-lying excited states
\cite{Alba:2009th,Alcaraz:2011tn,Berganza:2011mh,Pizorn:2012aut,Essler:2012rai,Berkovits:2013mii,%
Taddia:2013kxu,Storms:2013wzf,Palmai:2014jqa,Calabrese:2014ntv,Molter2014Bound,Taddia:2016dbm,%
Castro-Alvaredo:2018dja,Castro-Alvaredo:2018bij,Murciano:2018cfp,Castro-Alvaredo:2019irt,Castro-Alvaredo:2019lmj,%
Jafarizadeh:2019xxc,Capizzi:2020jed,You:2020osa,Haque:2020ewo,Zhang:2020ouz,Angel-Ramelli:2020xvd,Zhang:2020vtc,Wybo:2020fiz,%
Zhang:2020dtd,Zhang:2020txb,Eisler:2021uzt} and the typical states in the middle of the spectrum \cite{Deutsch:2012vkm,Santos:2012rss,Beugeling:2015plv,Garrison:2015lva} using exact and numerical calculations. See also \cite{Vidmar:2017uux,Vidmar:2017pak,Huang:2019dxk,Vidmar:2018rqk,LeBlond:2019eoe,Lydzba:2020qfx} for the behavior of the average entanglement entropy of the excited states.

When the system is integrable or in general when there is no particle production one can consider quasiparticle excitations whose entanglement shows a remarkable universal property which has a natural qubit interpretation \cite{Castro-Alvaredo:2018dja,Castro-Alvaredo:2018bij}, see also \cite{Pizorn:2012aut,Berkovits:2013mii,Molter2014Bound} for earlier similar conclusions.
In \cite{Castro-Alvaredo:2018dja,Castro-Alvaredo:2018bij} it was argued that the entanglement entropy for the quasiparticle excited  states  composed  of  finite  numbers  of quasiparticles with finite De Broglie wavelengths or finite intrinsic correlation length is largely independent  of  the  momenta  and  masses  of  the  excitations,  and  of  the  geometry,  dimension  and connectedness  of  the  entanglement  region. Using exact calculations on free fermions and bosons, i.e.\ Hamiltonian being equal to the number operator, these results were later confirmed and generalized in \cite{Zhang:2020ouz,Zhang:2020vtc,Zhang:2020dtd,Zhang:2020txb}. In particular it was shown that for the excited states with more than one mode excited in the most generic circumstances apart from the universal terms there are extra terms.
Especially, these additional terms have nontrivial dependence on the momenta of the excited quasiparticles.
These terms are non-negligible when the momentum difference of at least a pair of distinct quasiparticles is small. It was  argued that these extra terms are also universal though in a weaker sense. In particular, it was shown numerically that similar equations are valid also for the excess entanglement entropy of the excited state in slightly gapped and critical models.

In this paper we consider a circular chain with $L$ sites and choose its subsystem $A=[1,\ell]$ with $\ell$ consecutive sites. We calculate the quasiparticle excited state entanglement entropy between $A$ and its complement $B=[\ell+1,L]$ in the scaling limit $L\to+\inf$ and $\ell\to+\inf$ with fixed ratio $x=\f{\ell}{L}$. We first make a conjecture that the results presented in \cite{Zhang:2020vtc,Zhang:2020dtd} are valid not only for non-coupled systems, i.e. the Hamiltonian being just the totally free number operator, but also for generic translational invariant free fermionic and bosonic chains as far as the extra momenta are big.
We also made new conjectures for the differences of excited state entanglement entropy.
These conjectures are supported with numerical calculations. Then we calculate exactly the entanglement in magnon excited states of interacting chains such as spin-1/2 XXX and XXZ chains. We prove that when finite number of magnons are excited, in the scaling limit, one can decompose the entanglement content into sum of finite number of contributions that are exactly equal to the ones coming from non-coupled fermionic and bosonic chains with particular momenta.
In other words, one can classify the entanglement content of also interacting models with respect to the entanglement of the excited states of the free fermionic and bosonic theories.

The paper is organized as follows:
In the next section, i.e.\ section~\ref{SectionFermion}, we review the entanglement content of the excited states of the number operator in fermionic chains. Then from numerical calculations we conjecture that these results are universal in the scaling limit for generic quadratic free fermionic systems.
In section \ref{SectionBoson} we make similar arguments for bosonic chains.
In section \ref{SectionXXX} and \ref{SectionXXZ} we first calculate exactly the entanglement content of a few magnon states in the spin-1/2 XXX and XXZ chains and then show that in the scaling limit for the general magnon state the entanglement content can be written as the sum of fermionic and bosonic terms introduced in sections \ref{SectionFermion} and \ref{SectionBoson}.
Finally in section \ref{SectionConclusion} we conclude the paper with discussions.

The paper is accompanied with numerous appendices which in that we explore the exact calculation of the entanglement in free fermions, free bosons and the XXX chain.
In appendix~\ref{appNOB} we give the procedure to calculate the R\'enyi and entanglement entropies for density matrices in a general nonorthonormal basis.
In appendices~\ref{appFer} and \ref{appBos} we show examples of the analytical calculations of the R\'enyi and von Neumann entropies from the subsystem mode method in the free fermionic and bosonic chains.
In appendix~\ref{appBS} we discuss the R\'enyi and entanglement entropies and their various limits in the double-magnon bound state of the XXX chain.
In particular there are a few novel results in those appendices including the analytical and numerical calculations of the entanglement entropy, beyond the R\'enyi entropy with integer index $n\geq2$, from the subsystem mode method in the free fermions and free bosons in appendices~\ref{appFer} and \ref{appBos} and the analytical expression of various limits of the R\'enyi and entanglement entropies in the two-magnon bound states of the XXX chain in appendix~\ref{appBS}.

\section{Quadratic fermionic chain} \label{SectionFermion}

In this section we first review briefly the quasiparticle excited states in the simplest free fermionic chain, i.e.\ Hamiltonian without couplings. Then based on numerical calculations we make some conjectures regarding entanglement content of generic translational invariant free fermions.

The Hamiltonian of the most general translational invariant (periodic) quadratic fermionic chain with time-reversal symmetry takes the form
\be \label{fermionicgeneric}
H = \sum_{j=1}^L \sum_{r=0}^{L-1}
\Big[ \bA_r a_j^{\dagger}a_{j+r}
    + \frac{1}{2}(\bB_r a_j^{\dagger}a_{j+r}^{\dagger}- \bB_r^* a_j a_{j+r})
\Big]
- \f{1}{2} L \bA_0,
\ee
with the local fermionic modes $a_j$, $a_j^\dag$ and the parameters $\bA_r^*=\bA_{L-r}$, $\bB_r=-\bB_{L-r}$.
It could be diagonalized in terms of the translational invariant quasiparticle modes $c_k$, $c_k^\dag$ in the form
\be
H = \sum_{k} \ve_k \Big( c_k^\dag c_k - \f12 \Big),
\ee
where $\ve_k\geq0$ is the energy of the mode $k$.

The simplest possible case one can consider for the above Hamiltonian is the number operator:
\be \label{fermionicchainN}
N =  \sum_{j=1}^L \Big( a_j^\dag a_j - \f12 \Big).
\ee
The above Hamiltonian, for example, can be considered as the extremely gapped (infinite transverse field) limit of the spin-1/2 XY chain. For the above Hamiltonian we write the quasiparticle modes as
\be \label{fermionckckdagpk}
c_k = \f{1}{\sr{L}}\sum_{j=1}^L\ep^{-\ii j p_k}a_j, ~~
c_k^\dag = \f{1}{\sr{L}}\sum_{j=1}^L\ep^{\ii j p_k}a_j^\dag, ~~
p_k = \f{2\pi k}{L}.
\ee
Then the Hamiltonian becomes
\be
N = \sum_k \Big( c_k^\dag c_k - \f12 \Big).
\ee
Here $p_k$ is the actual momentum and $k$ is the total number of waves, which is an integer and a half-integer depending on the boundary conditions.
Note that $p_k\cong p_k+2\pi$ and $k\cong k+L$.
In this paper we will just call $p_k$ actual momentum and call $k$ momentum.
We only consider the case that $L$ is an even integer and the states in the Ramond sector, i.e.\ periodic boundary conditions for the spinless fermions $a_{L+1}=a_1$, $a_{L+1}^\dag=a_{1}^\dag$, and so we have the integer momenta
\be
k=1-\f{L}{2},\cdots,-1,0,1,\cdots,\f{L}{2}-1,\f{L}{2}.
\ee
The ground state $|G\rag$ is annihilated by all the lowering operators $c_k$
\be \label{fermionGrag}
c_k|G\rag=0,{\rm{~for~all~}} k,
\ee
and the general excited state $|K\rag$ is generated by applying the raising operators $c_k^\dag$ on the ground state
\be \label{fermionKrag}
|K\rag=|k_1\cdots k_r\rag=c_{k_1}^\dag\cdots c_{k_r}^\dag|G\rag.
\ee
The number of excited quasiparticles is $|K|=r$.

The entanglement entropy of the above states were studied comprehensively in \cite{Zhang:2020vtc,Zhang:2020dtd} using a technique called the subsystem mode method, see appendix \ref{appFer} for the explanation of the method and some extra new details and refinements that are based on the calculation of entropies in nonorthogonal basis which is explained in appendix \ref{appNOB}. The final result can be written as
\bea \label{entanglement-free-fermionI}
&& S_{A,K}^{(n),\fer} = - \f{1}{n-1}\log\tr\cR_{A,K}^n, \\
\label{entanglement-free-fermionII}
&& S_{A,K}^\fer = - \tr(\cR_{A,K}\log\cR_{A,K}).
\eea
where the exact form of the $2^{|K|}\times2^{|K|}$ matrix $\cR_{A,K}$  can be found in appendix~\ref{appSubsystem}. They are consistent with the formulas (\ref{FermionRenyi}) and (\ref{FermionvonNeumann}) coming from the correlation matrix method.
The above formulas (\ref{entanglement-free-fermionI}) and (\ref{entanglement-free-fermionII}) are quite efficient to calculate analytically and numerically the R\'enyi and entanglement entropies when finite number of modes are excited.
Some examples are provided in appendix \ref{appExamples}.

In the scaling limit, i.e.\ $L\to+\inf$ and $\ell\to+\inf$ with fixed ratio $x=\f{\ell}{L}$, the equations (\ref{entanglement-free-fermionI}) and (\ref{entanglement-free-fermionII}), or equivalently (\ref{FermionRenyi}) and (\ref{FermionvonNeumann}), can be calculated easily and they show a remarkable universal property \cite{Zhang:2020vtc,Zhang:2020dtd}.
The numerical results in the scaling limit of XY and XX chains suggest that the excess R\'enyi and entanglement entropies take the forms
\bea
&& S_{A,K}^{(n),\fer} - S_{A,G}^{(n),\fer} = - \f{1}{n-1}\log\tr\cR_{A,K}^n, \label{SAKnfermSAGnfer} \\
&& S_{A,K}^\fer - S_{A,G}^\fer = - \tr(\cR_{A,K}\log\cR_{A,K}), \label{SAKfermSAGfer}
\eea
as long as the momenta in the set $K$ satisfy%
\footnote{In \cite{Zhang:2020vtc,Zhang:2020dtd}, the proposed condition is \cite{Castro-Alvaredo:2018dja,Castro-Alvaredo:2018bij}
\[
\min\Big( \f1\D, \max_{k\in K}\f{L}{|k|} \Big) \ll \min( \ell, L-\ell ),
\]
where $\D$ is the gap of the model and the size of excited quasiparticle with momentum $k$ is represented as $\f{L}{|k|}$. From the example of the XX chain, now we think it is better to represent the size of excited quasiparticle with momentum $k$ as $\f{1}{\ve_k}$. The condition is modified as
\[
\min\Big( \f1\D, \max_{k\in K}\f{1}{\ve_k} \Big) \ll \min( \ell, L-\ell ).
\]
Supposing that $\ve_k$ is a continuous function of $k$, we have in the scaling limit
\[ \D=\min_{{\rm all~}k} \ve_k. \]
We obtain
\[
\min\Big( \max_{{\rm all~}k} \f{1}{\ve_k}, \max_{k\in K}\f{1}{\ve_k} \Big)
\ll
\min( \ell, L-\ell ),
\]
which is just the new condition (\ref{conditionNew}).
The physical meaning of the condition used in \cite{Castro-Alvaredo:2018dja,Castro-Alvaredo:2018bij,Zhang:2020vtc,Zhang:2020dtd} is that either the correlation length of the model or the maximal size of the excited quasiparticles is much smaller than the sizes of the subsystem and its complement. Using $\f{1}{\ve_k}$, instead of $\f{L}{|k|}$, to represent the size of the quasiparticle with momentum $k$, we see that the correlation length of the model is always larger than or equal to the maximal size of the excited quasiparticles, and we only need to impose the condition for the maximal size of the excited quasiparticles.}%
\be \label{conditionNew}
\max_{k\in K} \f{1}{\ve_k} \ll \min( \ell, L-\ell ).
\ee
We use $\f{1}{\ve_k}$ to represent the size of the quasiparticle with momentum $k$.
The above condition is just that all the sizes of excited quasiparticles are much smaller than the sizes of the subsystem $A$ and its complement $B$.
Note that with the new condition (\ref{conditionNew}), we no longer require that the gap of the model is large or each excited quasiparticle has a large momentum $k$. Instead, we only require that each excited quasiparticle has a large energy $\ve_k$.

The results in the previous paragraph could be further generalized.
Suppose we have two states $|K'\rag=|k'_1,\cdots,k'_{r'}\rag$ and $|K\cup K'\rag=|k_1,\cdots,k_r,k'_1,\cdots,k'_{r'}\rag$ in a general translational invariant fermionic chain of the Hamiltonian (\ref{fermionicgeneric}).
If the momenta in the set $K\cup K'$ satisfies (\ref{conditionNew}), from (\ref{SAKnfermSAGnfer}) and (\ref{SAKfermSAGfer}) we get
\bea
&& \hspace{-15mm}
   S_{A,K'}^{(n),\fer} - S_{A,G}^{(n),\fer} = - \f{1}{n-1}\log\tr\cR_{A,K'}^n, ~~
   S_{A,K'}^\fer - S_{A,G}^\fer = - \tr(\cR_{A,K'}\log\cR_{A,K'}),
   \label{SAKpnfermSAGnferSAKpfermSAGfer}\\
&& \hspace{-15mm}
   S_{A,K\cup K'}^{(n),\fer} - S_{A,G}^{(n),\fer} = - \f{1}{n-1}\log\tr\cR_{A,K\cup K'}^n, ~~
   S_{A,K\cup K'}^\fer - S_{A,G}^\fer = - \tr(\cR_{A,K\cup K'}\log\cR_{A,K\cup K'}). \label{SAKcupKpnfermSAGnferSAKcupKpfermSAGfer}
\eea
If the sets $K$ and $K'$ have large momentum differences in the scaling limit $L\to+\inf$, by which we mean
\be \label{kpmktopinf}
| k' - k | \to + \inf, \textrm{~for~all~} k' \in K' \textrm{~and~all~} k \in K,
\ee
the quasiparticles in $K$ and $K'$ would decoherent and contribute independently to the entanglement
\be
\cR_{A,K\cup K'} = \cR_{A,K} \otimes\cR_{A,K'}.
\ee
Then we get the differences of the R\'enyi and entanglement entropies
\bea
&& S_{A,K\cup K'}^{(n),\fer} - S_{A,K'}^{(n),\fer} = - \f{1}{n-1}\log\tr\cR_{A,K}^n, \label{SAKcupKpnfermSAKpnfer} \\
&& S_{A,K\cup K'}^\fer - S_{A,K'}^\fer = - \tr(\cR_{A,K}\log\cR_{A,K}). \label{SAKcupKpfermSAKpfer}
\eea
This is just a corollary of (\ref{SAKnfermSAGnfer}) and (\ref{SAKfermSAGfer}).
We emphasize that the above formulas (\ref{SAKcupKpnfermSAKpnfer}) and (\ref{SAKcupKpfermSAKpfer}) are derived under the condition:
\begin{itemize}
  \item The momenta in the set $K\cup K'$ satisfy (\ref{conditionNew});
  \item The sets $K$ and $K'$ have large momentum differences in the scaling limit $L\to+\inf$, i.e.\ (\ref{kpmktopinf}).
\end{itemize}

\begin{figure}[t]
  \centering
  \includegraphics[height=0.64\textwidth]{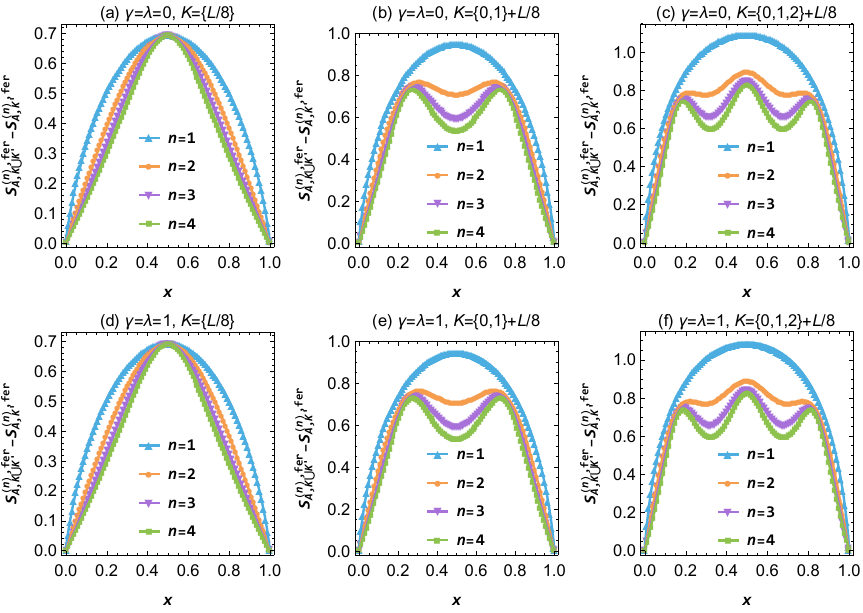}\\
  \caption{The differences of the R\'enyi and entanglement entropies $S_{A,K\cup K'}^{(n),\fer}-S_{A,K'}^{(n),\fer}$ with $n=1,2,3,4$ in the fermionic chain with the Hamiltonian (\ref{fermionicchainwithlamgam}).
  For the results from the correlation matrix method (symbols), i.e.\ LHS of (\ref{SAKcupKpnfermSAKpnfer}) and (\ref{SAKcupKpfermSAKpfer}), we have set $K'=\{1,2,\f{L}{4},\f{L}{4}+1\}$ and $L=1024$.
  For the results from the subsystem mode method (lines), i.e.\ RHS of (\ref{SAKcupKpnfermSAKpnfer}) and (\ref{SAKcupKpfermSAKpfer}), we have set $L=+\inf$.
  In panels (a) and (d), the analytical results from the subsystem mode method are (\ref{FermionSAkn}) and (\ref{FermionSAk}). In panels (b) and (e), the analytical results from the subsystem mode method are (\ref{FermionSAk1k2n}) and (\ref{FermionSAk1k2}) with $(k_1,k_2)=(0,1)$. In panels (c) and (f), the lines are just the R\'enyi and entanglement entropies $S_{A,012}^{(n),\fer}$ in the free fermionic chains.}
  \label{FigureFermionDeltaSAn}
\end{figure}

We note here that the excited state $|K\rag$ of the Hamiltonian (\ref{fermionicgeneric}) with specific $(\bA_r,\bB_r)$ can be actually the ground state $|G'\rag$ of another Hamiltonian with a new $(\bA'_r,\bB'_r)$, for some examples see \cite{Jafarizadeh:2019xxc}. The prime example
is the ground state of the XX chain which is an excited state of the number operator $N$ (\ref{fermionicchainN}) as the Hamiltonian.
Then one is tempted to guess that in the scaling limit as far as there is a gap between the two states then the equations (\ref{entanglement-free-fermionI}) and (\ref{entanglement-free-fermionII}) should work.
We conjecture that the formulas (\ref{SAKcupKpnfermSAKpnfer}) and (\ref{SAKcupKpfermSAKpfer}) are still valid under a weaker condition:
\begin{itemize}
  \item The momenta in the set $K$ satisfy (\ref{conditionNew});
  \item The sets $K$ and $K'$ have large momentum differences in the scaling limit $L\to+\inf$, i.e.\ (\ref{kpmktopinf}).
\end{itemize}
In particular, the set $K'$ may not satisfy (\ref{conditionNew}), and we do not have (\ref{SAKpnfermSAGnferSAKpfermSAGfer}) and (\ref{SAKcupKpnfermSAGnferSAKcupKpfermSAGfer}), but we still have (\ref{SAKcupKpnfermSAKpnfer}) and (\ref{SAKcupKpfermSAKpfer}).
Moreover, the set $K'$ could possibly has an infinite number of elements in the scaling limit.
This conjecture is beyond the results in \cite{Zhang:2020vtc,Zhang:2020dtd}.
To support the conjecture, in figure~\ref{FigureFermionDeltaSAn}, we show examples of the differences of the R\'enyi and entanglement entropies in the fermionic chain with Hamiltonian
\be \label{fermionicchainwithlamgam}
H = \sum_{j=1}^L \Big[ \lam \Big( a_j^\dag a_j - \f12 \Big) - \f12 ( a_j^\dag a_{j+1} + a_{j+1}^\dag a_j ) - \f{\g}{2} ( a_j^\dag a_{j+1}^\dag + a_{j+1} a_j ) \Big],
\ee
which is in fact the Jordan-Wigner transformation of the spin-1/2 XY chain.

\begin{figure}[t]
  \centering
  \includegraphics[height=0.64\textwidth]{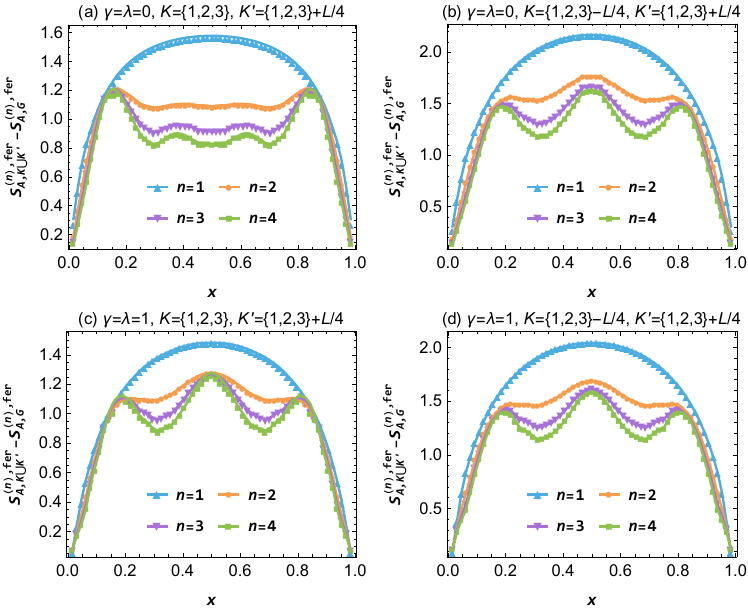}\\
  \caption{Check of the conjecture (\ref{SAK1cupK2nfer}) with the condition (\ref{kpmktopinf}) in the fermionic chain with the Hamiltonian (\ref{fermionicchainwithlamgam}).
  All the results are calculated numerically from the correlation matrix method with $L=64$.
  The LHS and RHS of (\ref{SAK1cupK2nfer}) are denoted by symbols and lines respectively.}
  \label{FigureFermionSAK1cupK2n}
\end{figure}

With the above results, we further conjecture that the formula
\be \label{SAK1cupK2nfer}
S_{A,K\cup K'}^{(n),\fer} - S_{A,G}^{(n),\fer} =
S_{A,K}^{(n),\fer} + S_{A,K'}^{(n),\fer} - 2 S_{A,G}^{(n),\fer},
\ee
is valid as long as the sets $K$ and $K'$ have large momentum differences in the scaling limit $L\to+\inf$, i.e.\ (\ref{kpmktopinf}).
We do not have other restrictions on the states or the model.
Note that in the formula (\ref{SAK1cupK2nfer}) the sets $K$ and $K'$ are on the same footing.
The conjecture (\ref{SAK1cupK2nfer}) has a simple physical meaning, and it asserts that the contributions to the entanglement from quasiparticles of large momentum differences decouple.
In fact, the conjecture (\ref{SAKcupKpnfermSAKpnfer}) and (\ref{SAKcupKpfermSAKpfer}) with the conditions (\ref{conditionNew}) and (\ref{kpmktopinf}) could be derived given the conjecture (\ref{entanglement-free-fermionI}) and (\ref{entanglement-free-fermionII}) with the condition (\ref{conditionNew}) plus the conjecture (\ref{SAK1cupK2nfer}) with the condition (\ref{kpmktopinf}).
We check the conjecture (\ref{SAK1cupK2nfer}) in the fermionic chain with the Hamiltonian (\ref{fermionicchainwithlamgam}) in figure~\ref{FigureFermionSAK1cupK2n}.
Note that in panels (b) and (c) of figure~\ref{FigureFermionSAK1cupK2n} the large energy condition (\ref{conditionNew}) for $K$ is broken and in panels (a) and (b) the large energy condition for $K'$ is also broken, while in all the panels the large momentum difference condition (\ref{kpmktopinf}) for $K$ and $K'$ is preserved.

\section{Quadratic bosonic chain} \label{SectionBoson}

This section is parallel to the previous section in many aspects.
We consider the Hamiltonian
\be \label{bosonicgeneric}
H = \sum_{j=1}^L \sum_{r=0}^{L-1}
\Big[ \bA_r a_j^{\dagger}a_{j+r}
    + \frac{1}{2}(\bB_r a_j^{\dagger}a_{j+r}^{\dagger} + \bB_r^* a_j a_{j+r})
\Big]
+ \f{1}{2} L \bA_0,
\ee
with the bosonic modes $a_j$, $a_j^\dag$ and the parameters $\bA_r^*=\bA_{L-r}$, $\bB_r=\bB_{L-r}$.

As before we first study the uncoupled Hamiltonian
\be \label{uncoupled-bosonicchain}
N = \sum_{j=1}^L \Big( a_j^\dag a_j + \f12 \Big),
\ee
and then make conjectures regarding more general Hamiltonians.
In this case a general excited state takes the form
\be
|K\rag=|k_1^{r_1}\cdots k_s^{r_s}\rag =
\f{( c^\dag_{k_1} )^{r_1} \cdots ( c^\dag_{k_s} )^{r_s}}{\sr{N_{r_1\cdots r_s}}} | G \rag,
\ee
with the quasiparticle modes $c_k$, $c_k^\dag$ defined the same as (\ref{fermionckckdagpk}), the ground state $| G \rag$ defined the same as (\ref{fermionGrag}), and the normalization factor
\be
N_K=N_{r_1\cdots r_s} = r_1!\cdots r_s!.
\ee

The subsystem mode method can be also used in this case, see appendix~\ref{appBos}, and this leads to the same equations of the R\'enyi and entanglement entropies as (\ref{entanglement-free-fermionI}) and (\ref{entanglement-free-fermionII})
\bea
&& S_{A,K}^{(n),\bos} = - \f{1}{n-1}\log\tr\cR_{A,K}^n, \label{entanglement-free-bosonI}\\
&& S_{A,K}^\bos = - \tr(\cR_{A,K}\log\cR_{A,K}), \label{entanglement-free-bosonII}
\eea
where the exact form of the matrix $\cR_{A,K}$ can be found in appendix \ref{appSubsystem2}. Since in this case the excited states are not Gaussian states, instead of the correlation matrix method, one has the wave function method \cite{Castro-Alvaredo:2018dja,Castro-Alvaredo:2018bij}, which leads to
a permanent formula for the R\'enyi entropy in the uncoupled bosonic chain with the Hamiltonian (\ref{uncoupled-bosonicchain}) \cite{Zhang:2020dtd}
\be \label{BosonRenyiPermanentFormulaText}
S_{A,K}^{(n),\bos} = - \f{1}{n-1} \log \f{\per \O_{A,K}^{(n)}}{N_K^n},
\ee
with the $n|K|\times n|K|$ matrix $\Omega_{A,K}^{(n)}$ is written in appendix~\ref{appwave-function}.
The formula of the R\'enyi entropy (\ref{BosonRenyiPermanentFormulaText}) is applicable only for integer $n\geq2$. In appendix \ref{appExamples2} we provide some examples regarding the RDM and the entanglement entropy of the states in which a few modes are excited. Some of the results are beyond the cases that were discussed in \cite{Zhang:2020vtc,Zhang:2020dtd}.

\begin{figure}[t]
  \centering
  \includegraphics[height=0.32\textwidth]{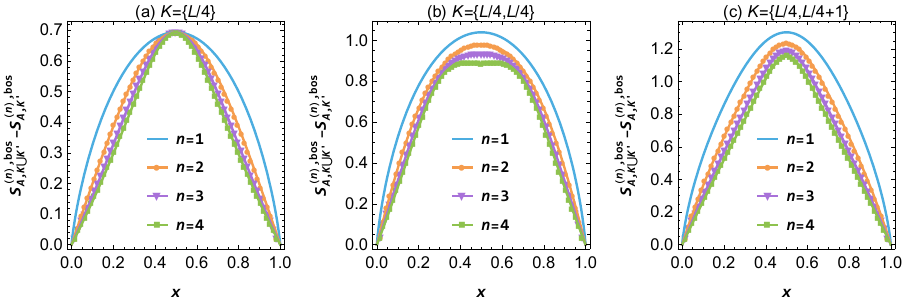}\\
  \caption{The differences of the R\'enyi and entanglement entropies $S_{A,K\cup K'}^{(n),\bos}-S_{A,K'}^{(n),\bos}$ with $n=1,2,3,4$ in the discrete Klein-Gordon field theory with mass $m$ (\ref{KGbosonhamiltonian}).
  For the results from the wave function method (symbols), i.e.\ LHS of (\ref{SAKcupKpnbosSAKpnbos}) and (\ref{SAKcupKpbosSAKpbos}), we have set $m=10^{-5}$, $K'=\{1\}$ and $L=64$.
  Note that we do not have the entanglement entropies, i.e.\ the R\'enyi entropies with $n=1$, from the wave function method.
  For the results from the subsystem mode method (lines), i.e.\ RHS of (\ref{SAKcupKpnbosSAKpnbos}) and (\ref{SAKcupKpbosSAKpbos}), we have set $L=+\inf$.
  In panel (a) the analytical results are (\ref{BosonSAkn}) and (\ref{BosonSAk}).
  In panel (b) the analytical results are (\ref{BosonSAk2n}) and (\ref{BosonSAk2}).
  In panel (c) the analytical results are (\ref{BosonSAk1k2n}) and (\ref{BosonSAk1k2}) with $(k_1,k_2)=(0,1)$.}
  \label{FigureBosonDeltaSAn}
\end{figure}

Numerical calculations done on the discrete Klein-Gordon field theory presented in \cite{Zhang:2020vtc,Zhang:2020dtd} suggest that, similar to the free fermion case, here too the above equations (\ref{entanglement-free-bosonI}) and (\ref{entanglement-free-bosonII}) give a good approximation also for the excess entropy in slightly gapped and even critical chains as far as the momenta $k$'s are large.
This makes it plausible to guess that the conjecture
\bea
&& S_{A,K\cup K'}^{(n),\bos} - S_{A,K'}^{(n),\bos} = - \f{1}{n-1}\log\tr\cR_{A,K}^n, \label{SAKcupKpnbosSAKpnbos} \\
&& S_{A,K\cup K'}^\bos - S_{A,K'}^\bos = - \tr(\cR_{A,K}\log\cR_{A,K}). \label{SAKcupKpbosSAKpbos}
\eea
might be also valid for free bosonic chains given the conditions
\begin{itemize}
  \item The momenta in the set $K$ satisfy (\ref{conditionNew});
  \item The sets $K$ and $K'$ have large momentum differences in the scaling limit $L\to+\inf$, i.e.\ (\ref{kpmktopinf}).
\end{itemize}
We make new numerical calculations on discrete Klein-Gordon field theory with mass $m$, i.e.\ the Hamiltonian
\be \label{KGbosonhamiltonian}
H = \sum_{j=1}^L \Big[
\Big(m+\f1m\Big) \Big( a^\dag_j a_j + \f12 \Big)
- \f1{2m}( a^\dag_j a_{j+1} + a_j a^\dag_{j+1} )
+\f1{2m} [ (a^\dag_j)^2 - a^\dag_j a^\dag_{j+1} + a_j^2 - a_j a_{j+1} ] \Big],
\ee
as shown in figure~\ref{FigureBosonDeltaSAn}, supporting this conjecture.
Note that in the figure there are no entanglement entropies from the wave function method.

\begin{figure}[t]
  \centering
  \includegraphics[height=0.32\textwidth]{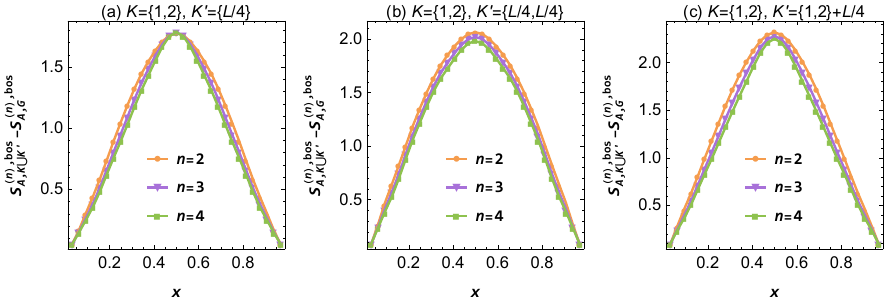}\\
  \caption{Check of the conjecture (\ref{SaK1cupK2nbosmSAGnbos}) with the condition (\ref{kpmktopinf}) in the discrete Klein-Gordon field theory (\ref{KGbosonhamiltonian}).
  All the results are calculated numerically from the correlation matrix method with $m=10^{-5}$, $L=32$.
  The LHS and RHS of (\ref{SaK1cupK2nbosmSAGnbos}) are represented by symbols and lines respectively.}
  \label{FigureBosonSAK1cupK2n}
\end{figure}

The same as that in the fermionic chain, we further conjecture that there is
\be \label{SaK1cupK2nbosmSAGnbos}
S_{A,K\cup K'}^{(n),\bos} - S_{A,G}^{(n),\bos} =
S_{A,K}^{(n),\bos} + S_{A,K'}^{(n),\bos} - 2 S_{A,G}^{(n),\bos},
\ee
with the condition (\ref{kpmktopinf}).
We check the conjecture in the discrete Klein-Gordon field theory (\ref{KGbosonhamiltonian}) in figure~\ref{FigureBosonSAK1cupK2n}.
In all the panels of figure~\ref{FigureBosonSAK1cupK2n} the large energy condition (\ref{conditionNew}) for $K$ is broken, while in all the panels the large momentum difference condition (\ref{kpmktopinf}) is preserved.

\section{XXX chain}\label{SectionXXX}

In this section we calculate exactly the R\'enyi and entanglement entropies in the magnon excited states in the XXX chain.
We only consider the states with finite numbers of magnons and mainly focus on the scattering states.
For the bound states see appendix~\ref{appBS}.
In the scaling limit, the magnon excited state R\'enyi and entanglement entropies turn out to be simply related to the R\'enyi and entanglement entropies in the fermionic and bosonic chains with the number operator as the Hamiltonian.
This shows remarkably that in the scaling limit the equations (\ref{entanglement-free-fermionI}) and (\ref{entanglement-free-fermionII}) for free fermions and the equations (\ref{entanglement-free-bosonI}) and (\ref{entanglement-free-bosonII}) for free bosons are still valid even beyond non-interacting free Hamiltonians.

\subsection{Magnon excited states}

We consider the spin-1/2 XXX chain in transverse field
\be
H = -\f14 \sum_{j=1}^L ( \s_j^x\s_{j+1}^x + \s_j^y\s_{j+1}^y + \s_j^z\s_{j+1}^z ) - \f{h}{2} \sum_{j=1}^L \s_j^z,
\ee
with total number of sites $L$ being four times of an integer and periodic boundary conditions for the Pauli matrices $\s_{L+1}^{x,y,z}=\s_1^{x,y,z}$.
We only consider the ferromagnetic phase with $h > 0$, so that the unique ground state is
\be \label{XXXG}
|G\rag = |\!\uparrow\uparrow\cdots\uparrow\rag,
\ee
and low energy excitations are magnons.

The XXX chain can be solved with the Bethe ansatz, for review see for example \cite{Gaudin:1983kpuCaux:2014uuq,Karbach:1998abi}.
A general state with $R$ magnons takes the form
\be \label{statepsi}
|\cI\rag = \f{1}{\sr{\cN}} \sum_{1\leq j_1 < \cdots <j_R\leq L} \cU_{j_1\cdots j_R} |j_1\cdots j_R\rag,
\ee
with the normalization factor
\be \label{statepsicN}
\cN = \sum_{1\leq j_1 < \cdots <j_R\leq L} | \cU_{j_1\cdots j_R} |^2.
\ee
The state $|j_1\cdots j_R\rag$ is the configuration that the spins on the sites $j_1,\cdots,j_R$ are flipped.
The ansatz for the wave function is
\be \label{statepsicUj1cdotsjR}
\cU_{j_1\cdots j_R} = \sum_{\cP \in \cS_R}
\exp\Big( \ii \sum_{i=1}^R j_i p_{\cP i}
        + \f{\ii}{2} \sum_{1\leq i < i' \leq R} \th_{\cP i \cP i'} \Big),
\ee
where the sum $\cP$ is over all the $R!$ elements of the permutation group $\cS_R$.
The phase shift $\th_{ii'}$ is determined by the actual momenta $p_i$, $p_{i'}$ through
\be
\ep^{\ii\th_{ii'}} = - \f{1+\ep^{\ii(p_i+p_{i'})}-2\ep^{\ii p_{i}}}
                         {1+\ep^{\ii(p_{i}+p_{i'})}-2\ep^{\ii p_{i'}}}.
\ee
There is $\th_{ii'}=-\th_{i'i}$.
The actual momenta $p_i$ are related to the Bethe numbers $I_i \in [0,L-1]$ as
\be
L p_i = 2 \pi I_i + \sum_{i'\neq i} \th_{ii'}.
\ee

In the XXX chain we use the actual momentum $p=\f{2\pi k}{L}$ and the momentum $k$ (which is actually the total wave number) with $p\cong p+2\pi$ and $k\cong k+L$.
We have
\be \label{XXXkifromIi}
k_i = I_i + \f{1}{2\pi}\sum_{i'\neq i} \th_{ii'}.
\ee
Note that the Bethe numbers $I_i$ are integers.
For general $L$, the momenta $k_i$ are not necessarily integers or half-integers, and $k_i$ can be complex numbers for bound states.
In this paper, we mainly focus on the scattering magnon excited states, for which all the momenta $k_i$ are real numbers.
We use Bethe quantum numbers of the excited magnons $\cI=\{I_1,I_2,\cdots,I_R\}$ to represent the magnon excited states in the XXX chain.

\subsection{Local mode method}

We have the subsystem $A=[1,\ell]$ and its complement $B=[\ell+1,L]$, and write the site index $j\in[1,L]$ as $j=a$ or $j=b$ with $a \in [1,\ell]$ and $b \in [\ell+1,L]$.
The general magnon excited state (\ref{statepsi}) is just
\be \label{statepsiX}
|\cI\rag =
\f{1}{\sr{\cN}}
\sum_{i=0}^R
\sum_{1\leq a_1 < \cdots <a_i\leq \ell,
\ell+1\leq b_{i+1} < \cdots <b_R\leq L}
\cU_{a_1\cdots a_i b_{i+1}\cdots b_R} |a_1\cdots a_i b_{i+1} \cdots b_R\rag.
\ee
For each $i=0,\cdots,R$, we define the new indices $\cI_i=\{a_1,\cdots,a_i\}$ and $\cJ_i=\{b_{i+1},\cdots,b_R\}$, which characterize the configurations of the subsystems $A$ and $B$ respectively, and write the tensor $\cU$ as a $C_\ell^i \times C_{L-\ell}^{R-i}$ matrix $\cU_i$ with entries
\be
[\cU_i]_{\cI_i\cJ_i} = \cU_{a_1\cdots a_i b_{i+1}\cdots b_R}.
\ee
We write the state (\ref{statepsiX}) as
\be
|\cI\rag = \f{1}{\sr{\cN}} \sum_{i=0}^R\sum_{\cI_i,\cJ_i}[\cU_i]_{\cI_i \cJ_i}|\cI_i \cJ_i\rag.
\ee
Note that the basis is orthonormal
\be
\lag \cI_i \cJ_i|I'_{i'} \cJ'_{i'}\rag = \d_{ii'}\d_{\cI_iI'_{i}}\d_{\cJ_i\cJ'_{i}}.
\ee
Then we get the RDM in an orthonormal basis
\be \label{rA}
\r_{A,\cI} = \sum_{i=0}^R \sum_{\cI_i,\cI'_i} [\cV_i]_{\cI_i\cI'_i} |\cI_i\rag \lag \cI'_i|,
\ee
with the definition of the $C_\ell^i \times C_\ell^i$ matrix
\be
\cV_i = \f{\cU_i \cU_i^\dag}{\cN}.
\ee
We get the R\'enyi entropy
\be \label{SAcInXXX}
S_{A,\cI}^{(n),\XXX} = - \f{1}{n-1} \log\Big( \sum_{i=0}^R \tr\cV_i^n \Big),
\ee
and entanglement entropy
\be  \label{SAcIXXX}
S_{A,\cI}^\XXX = - \sum_{i=0}^R \tr [\cV_i\log\cV_i].
\ee
Note that $\cV_i$ is well-defined only for $i$ in the range
\be
\max(0,R+\ell-L) \leq i \leq \min(\ell,R),
\ee
and we define $\cV_i=0$ for other values of $i$.
The local mode method is not so efficient, and only applies to the states with the excitations of a few magnons in the XXX chain with a not so large number of sties $L$.

\subsection{Ground state}

The ground state (\ref{XXXG}) is a direct product state
\be
|G\rag = |G_A\rag|G_B\rag.
\ee
The R\'enyi and entanglement entropies in the ground state are just vanishing
\be
S_{A,G}^{(n),\XXX}=S_{A,G}^\XXX=0.
\ee

\subsection{Single-magnon states}

When only one particle is excited, there is the state
\be
|I\rag = \f{1}{\sqrt{L}} \sum_{j=1}^L \ep^{\f{2\pi\ii j I}{L}} |j\rag.
\ee
We have the Bethe quantum number $I\in[0,L-1]$ with $I\cong I+L$.
There are actual momentum $p$ and momentum $k$ with the relation $p=\f{2\pi k}{L}=\f{2\pi I}{L}$.
The R\'enyi and entanglement entropies are the same as the R\'enyi and entanglement entropies in the Hamiltonian $N$, see (\ref{FermionSAkn}), (\ref{FermionSAk}), (\ref{BosonSAkn}) and (\ref{BosonSAk}), i.e.
\bea
&& S_{A,I}^{(n),\XXX} = -\f{1}{n-1}\log[x^n+(1-x)^n], \\
&& S_{A,I}^\XXX = - x\log x - (1-x)\log(1-x).
\eea
These expressions have been obtained in \cite{Pizorn:2012aut,Berkovits:2013mii,Molter2014Bound,Castro-Alvaredo:2018dja,%
Castro-Alvaredo:2018bij}.
As $S_{A,k}^{(n),\fer}=S_{A,k}^{(n),\bos}=S_{A,I}^{(n),\XXX}$ for any $k$ and any $I$, and we will just write all the single-particle state R\'enyi and entanglement entropies as $S_{A,k}^{(n)}$.

\subsection{Double-magnon states}

The double-magnon states can be scattering states or bound states.
We focus on the scattering states in this subsection, and we discuss the bound states in appendix~\ref{appBS}.
The calculations in this subsection have overlaps with \cite{Molter2014Bound,Castro-Alvaredo:2018dja}.

We consider the states
\be \label{XXXI1I2}
|I_1I_2\rag = \f{1}{\sr{\cN}} \sum_{1\leq j_1 < j_2 \leq L} \cU_{j_1j_2}|j_1j_2\rag,
\ee
with Bethe numbers $I_1,I_2$,
\be  \label{cUj1j2}
\cU_{j_1j_2} = \ep^{\ii(j_1p_1+j_2p_2+\f12\th)} + \ep^{\ii(j_1p_2+j_2p_1-\f12\th)},
\ee
and the normalization factor
\be \label{cN}
\cN=\sum_{1\leq j_1 < j_2 \leq L} |\cU_{j_1j_2}|^2.
\ee
Generally there are $0\leq I_1\leq I_2\leq L-1$.
The two magnons have physical momenta $p_1$, $p_2$ and momenta $k_1$, $k_2$
\be
p_1 = \f{2\pi k_1}{L}, ~~
p_2 = \f{2\pi k_2}{L}.
\ee
Note that $k_1,k_2$ may not necessarily be integers and may be possibly complex numbers.
The total actual momentum, total momentum, and total Bethe number of the state are
\be
p=p_1+p_2, ~~ k=k_1+k_2, ~~ I=I_1+I_2,
\ee
with
\be
p = \f{2\pi k}{L}, ~~ k = I.
\ee
The total Bethe number $I$ is an integer in the range $[0,2L-2]$.
The angle $\th$ is determined by the equation
\be \label{Betheequation}
\ep^{\ii\th} = - \f{1+\ep^{\ii(p_1+p_2)}-2\ep^{\ii p_1}}{1+\ep^{\ii(p_1+p_2)}-2\ep^{\ii p_2}}.
\ee

To the equation (\ref{Betheequation}), there are three classes of solutions \cite{Karbach:1998abi}.
The first two classes are scattering states, and the third class are bound states.
The class I solutions have $I_1=0$, $I_2=0,1,\cdots,L-1$, and $\th=0$.
The class I solutions are just excitations of hardcore bosons.
The class II solutions have $0<I_1<I_2\leq L-1$.
Most of the class II solutions have $I_1\leq I_2-2$, and there are also cases with $I_1= I_2-1$.
The class III solutions have $I_1=I_2$ or $I_1=I_2-1$.
We rename all the two-magnon solutions.
We call the special class I solutions with $I_1=I_2=0$ case I solutions.
We call both the class I solutions with $I_2\neq0$ and all class II solutions case II solutions.
We call all class III solutions case III solutions.
We discuss the R\'enyi and entanglement entropies in case I and case II solutions in the following two subsections, and we discuss the R\'enyi and entanglement entropies in case III solutions in appendix~\ref{appBS}.

\subsubsection{Case I solution}

For the case I solution, there are
\be
p_1=I_1=k_1=p_2=I_2=k_2=\th=0.
\ee
The state is
\be
|00\rag = \sr{\f{2}{L(L-1)}} \sum_{1\leq j_1<j_2\leq L} |j_1j_2\rag,
\ee
In the orthonormal basis of the states in the subsystem $A$
\be
|\psi_0\rag = |G_A\rag, ~~
|\psi_1\rag = \f{1}{\sr{\ell}}\sum_{j=1}^\ell|j\rag, ~~
|\psi_2\rag = \sr{\f{2}{\ell(\ell-1)}}\sum_{1\leq j_1<j_2\leq \ell}|j_1j_2\rag,
\ee
we have the diagonal RDM
\be
\r_{A,00}=\f{1}{L(L-1)}  \big[
   (L-\ell)(L-\ell-1) |\psi_0\rag\lag\psi_0|
 + 2\ell(L-\ell) |\psi_1\rag\lag\psi_1|
 +\ell(\ell-1) |\psi_2\rag\lag\psi_2|
\big].
\ee
We obtain the exact R\'enyi and entanglement entropies
\bea
&& S_{A,00}^{(n),\XXX} = - \f{1}{n-1} \log \f{[\ell(\ell-1)]^n+[2\ell(L-\ell)]^n+[(L-\ell)(L-\ell-1)]^n}{[L(L-1)]^n}, \\
&& S_{A,00}^\XXX = - \f{\ell(\ell-1)}{L(L-1)} \log \f{\ell(\ell-1)}{L(L-1)}
              - \f{2\ell(L-\ell)}{L(L-1)} \log \f{2\ell(L-\ell)}{L(L-1)} \nn\\
&& \phantom{S_{A,00} =}
              - \f{(L-\ell)(L-\ell-1)}{L(L-1)} \log \f{(L-\ell)(L-\ell-1)}{L(L-1)}.
\eea
In the scaling limit $L\to+\inf$ and $\ell\to+\inf$ with fixed ratio $x=\f{\ell}{L}$, the R\'enyi and entanglement entropies become
\be
S_{A,00}^{(n),\XXX} = S_{A,0^2}^{(n),\bos}, ~~
S_{A,00}^\XXX = S_{A,0^2}^{\bos},
\ee
with $S_{A,0^2}^{(n),\bos}$ and $S_{A,0^2}^{\bos}$ being the results (\ref{BosonSAk2n}) and (\ref{BosonSAk2}) in the bosonic chain.

\subsubsection{Case II solutions}

For case II solutions, there are
\be
p_1=\f{2\pi I_1+\th}{L}, ~~
p_2=\f{2\pi I_2-\th}{L}.
\ee
Note that
\be \label{k1k2fromI1I2theta}
k_1=I_1+\f{\th}{2\pi}, ~~
k_2=I_2-\f{\th}{2\pi}.
\ee
The angle $\th \in [-\pi,\pi]$ is a real solution to the equation (\ref{Betheequation}).
We use the nonorthonormal basis
\bea \label{XXXphi0123}
&& |\phi_0\rag = |G_A\rag, \nn\\
&& |\phi_1\rag = \f{1}{\sr{L}} \sum_{j=1}^\ell \ep^{\ii j p_1} |j\rag, \nn\\
&& |\phi_2\rag = \f{1}{\sr{L}} \sum_{j=1}^\ell \ep^{\ii j p_2} |j\rag, \nn\\
&& |\phi_3\rag = \f{1}{L} \sum_{1 \leq j_1 < j_2 \leq  \ell} \cU_{j_1j_2} |j_1j_2\rag,
\eea
with the definition of $\cU_{j_1j_2}$ (\ref{cUj1j2}).
For the state $|I_1I_2\rag$ there are the $4\times4$ matrices
\bea
&& \cP_{A,I_1I_2} = \f{L^2}{\cN}
\lt(\ba{cccc}
\cB &                           &                      & \\
    & 1-x                       & \ep^{\ii\th} \b_{12} & \\
    & \ep^{-\ii\th} \bar\b_{12} & 1-x                  & \\
    &                           &                      & 1
\ea\rt), \\
&& \cQ_{A,I_1I_2} =
\lt(\ba{cccc}
1 &             &         & \\
  & x           & \a_{12} & \\
  & \bar\a_{12} & x       & \\
  &             &         & \cA
\ea\rt), \\
&& \cR_{A,I_1I_2} = \f{L^2}{\cN}
\lt(\ba{cccc}
\cB &                                            &                                        & \\
    & x(1-x)+\ep^{\ii\th}\bar\a_{12}\b_{12}      & (1-x)\a_{12}+x\ep^{\ii\th}\b_{12}      & \\
    & (1-x)\bar\a_{12}+x\ep^{-\ii\th}\bar\b_{12} & x(1-x)+\ep^{-\ii\th}\a_{12}\bar\b_{12} & \\
    &                                            &                                        & \cA
\ea\rt), \label{XXXcRAI1I2}
\eea
with the definitions $\a_{12}\equiv \a_{k_1-k_2}$, $\b_{12}\equiv \b_{k_1-k_2}$, where $\a_k$ and $\b_k$ are defined as (\ref{alphak}) and (\ref{betak}), and
\bea
&& \cN = L(L-1) + \f{L\cos(\th-p_{12})-(L-1)\cos\th-\cos(\th-L p_{12})}{1-\cos p_{12}}, \nn\\
&& \cA = \f{1}{L^2} \Big[ \ell(\ell-1) + \f{\ell\cos(\th-p_{12})-(\ell-1)\cos\th-\cos(\th-\ell p_{12})}{1-\cos p_{12}} \Big], \\
&& \cB = \f{1}{L^2} \Big[ (L-\ell)(L-\ell-1) + \f{(L-\ell)\cos(\th-p_{12})-(L-\ell-1)\cos\th-\cos[\th-(L-\ell) p_{12}]}{1-\cos p_{12}} \Big], \nn
\eea
as well as $p_{12}\equiv p_1-p_2$. The matrix $\cR_{A,I_1I_2}$ (\ref{XXXcRAI1I2}) has four eigenvalues $\n_i$ with $i=1,2,3,4$ which are
\bea
&& \n_1 = \f{L^2\cA}{\cN}, ~~
   \n_2 = \f{L^2\cB}{\cN}, \nn\\
&& \n_{3/4} = \f{1}{\cN}
              \Big\{
                  \ell(L-\ell)
                + \cos\Big( \th-\f{L p_{12}}{2} \Big)
                  \f{\sin\f{\ell p_{12}}{2}\sin\f{(L-\ell)p_{12}}{2}}{\sin^2\f{p_{12}}{2}}
                  \pm \Big[
                          \ell^2 \f{\sin^2\f{(L-\ell)p_{12}}{2}}{\sin^2\f{p_{12}}{2}} \nn\\
&& \phantom{\n_{3/4} =}
                        + (L-\ell)^2\f{\sin^2\f{\ell p_{12}}{2}}{\sin^2\f{p_{12}}{2}}
                        + 2 \ell(L-\ell)
                            \cos\Big( \th-\f{L p_{12}}{2} \Big)
                            \f{\sin\f{\ell p_{12}}{2}\sin\f{(L-\ell)p_{12}}{2}}{\sin^2\f{p_{12}}{2}} \nn\\
&& \phantom{\n_{3/4} =}
                        - \sin^2\Big( \th-\f{L p_{12}}{2} \Big)
                          \f{\sin^2\f{\ell p_{12}}{2}\sin^2\f{(L-\ell)p_{12}}{2}}
                            {\sin^4\f{p_{12}}{2}}
                      \Big]^{1/2}
              \Big\}.
\eea
The R\'enyi and entanglement entropies are
\bea
&& S_{A,I_1I_2}^{(n),\XXX} = - \f{1}{n-1} \log \Big( \sum_{i=1}^4 \n_i^n \Big), \label{SAI1I2nXXX}\\
&& S_{A,I_1I_2}^\XXX = - \sum_{i=1}^4 \n_i \log \n_i \label{SAI1I2XXX}.
\eea

\begin{figure}[t]
  \centering
  \includegraphics[height=0.31\textwidth]{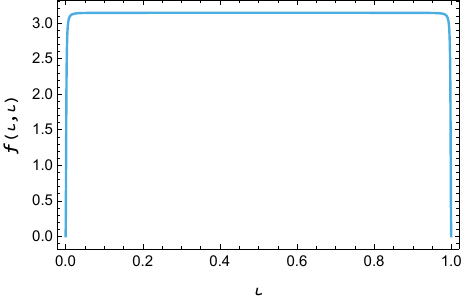}\\
  \caption{The shift angle $\th=f(\io,\io)$ with equal $\io_1=\io_2=\io$ in the scaling limit of the XXX chain. There is $f(0,0)=f(1,1)=0$, and there is $f(\io,\io)=\pi$ for $0<\io<1$.}
  \label{FigureXXXtheta}
\end{figure}

We are interested in the R\'enyi and entanglement entropies (\ref{SAI1I2nXXX}) and (\ref{SAI1I2XXX}) in the scaling limit $L\to+\inf$ and $\ell\to+\inf$ with fixed $x=\f{\ell}{L}$.
The shift angle becomes
\be \label{XXXBetheAngle}
\lim_{L\to+\inf} \theta = f(\io_1,\io_2),
\ee
with the scaled Bethe numbers
\be \label{ratiosi1i2}
\io_1 = \lim_{L\to+\inf} \f{I_1}{L}, ~~ \io_2 = \lim_{L\to+\inf} \f{I_2}{L},
\ee
and the function
\be \label{fio1io2}
f(\io_1,\io_2)=
\f1\ii \log \Big( - \f{1+\ep^{2\pi\ii(\io_1+\io_2)}- 2\ep^{2\pi\ii \io_1}}
    {1+\ep^{2\pi\ii(\io_1+\io_2)}- 2\ep^{2\pi\ii \io_2}} \Big).
\ee
There is $0\leq\io_1\leq\io_2\leq1$.
As the Bethe numbers satisfy $I\cong I+L$, we have the scaled Bethe numbers satisfying $\io\cong\io+1$.
When $\io_1=\io_2=\io$, it is understood that
\be
f(\io,\io) = \lim_{\e\to0^+} f(\io-\e,\io+\e).
\ee
We plot the shift angle $\th=f(\io,\io)$ in figure~\ref{FigureXXXtheta}.

For the cases with $\io_1=\io_2=0$, or $\io_1=\io_2=1$, or $\io_1=0$ and $\io_2=1$, there is the shift angle $\th=0$ and the states are excitations of two hardcore bosons.
In the scaling limit, the R\'enyi and entanglement entropies (\ref{SAI1I2nXXX}) and (\ref{SAI1I2XXX}) become
\be \label{SAI1I2nXXXcaseIIa}
S_{A,I_1I_2}^{(n),\XXX} = S_{A,k_1k_2}^{(n),\bos}, ~~
S_{A,I_1I_2}^\XXX = S_{A,k_1k_2}^{\bos}, ~~
k_1=I_1, ~ k_2 = I_2,
\ee
where $S_{A,k_1k_2}^{(n),\bos}$ and $S_{A,k_1k_2}^{\bos}$ are just the double-particle R\'enyi and entanglement entropies in the extremely gapped bosonic chain (\ref{BosonSAk1k2n}) and (\ref{BosonSAk1k2}).

For the cases with $\io_1=\io_2=\io\in(0,1)$, there is shift angle $\th=\pi$ and the states are excitations of two fermions, we have the R\'enyi and entanglement entropies
\be \label{SAI1I2nXXXcaseIIb}
S_{A,I_1I_2}^{(n),\XXX} = S_{A,k_1k_2}^{(n),\fer}, ~~
S_{A,I_1I_2}^\XXX = S_{A,k_1k_2}^{\fer}, ~~
k_1=I_1+\f12, ~ k_2 = I_2-\f12,
\ee
where $S_{A,k_1k_2}^{(n),\fer}$ and $S_{A,k_1k_2}^{\fer}$ are just the double-particle R\'enyi and entanglement entropies in the extremely gapped bosonic chain (\ref{FermionSAk1k2n}) and (\ref{FermionSAk1k2}).

For the remaining cases with $0 \leq \io_1<\io_2 \leq 1$ excluding the case with $\io_1=0$, $\io_2=1$, there is the shift angle $\th \in [0,\pi)$, whose explicit value is not important to us.
The two magnons have a large momentum difference
\be
|k_1-k_2| \to \Big| L(\io_1-\io_2)+\f{\theta}{\pi}\Big| \to + \inf,
\ee
and their contributions to the R\'enyi and entanglement entropies are independent.
We get the R\'enyi and entanglement entropies
\bea \label{SAI1I2nXXXcaseIIc}
&& S_{A,I_1I_2}^{(n),\XXX} = - \f{2}{n-1} \log[ x^n + (1-x)^n ]
                      = S_{A,I_1}^{(n),\XXX}+S_{A,I_2}^{(n),\XXX}
                      = 2 S_{A,k}^{(n)},  \\
&& S_{A,I_1I_2}^\XXX = 2 [ - x\log x - (1-x) \log (1-x) ]
                = S_{A,I_1}^\XXX + S_{A,I_2}^\XXX
                = 2 S_{A,k}.
\eea
They are just the universal R\'enyi and entanglement entropies obtained in \cite{Castro-Alvaredo:2018dja,Castro-Alvaredo:2018bij}.

\begin{figure}[t]
  \centering
  \includegraphics[height=0.31\textwidth]{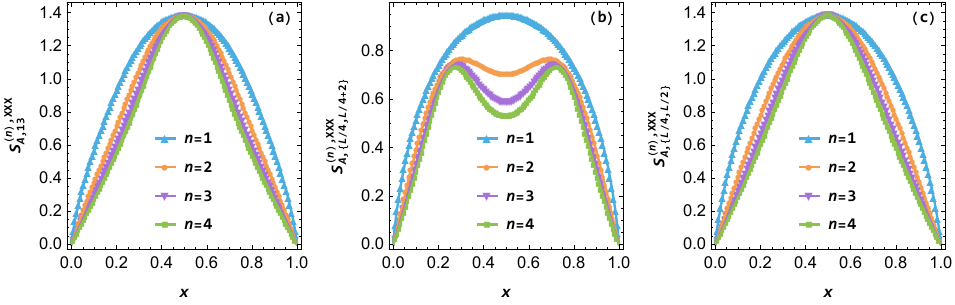}\\
  \caption{Exact results of the R\'enyi and entanglement entropies $S_{A,I_1I_2}^{(n),\XXX}$ with $n=1,2,3,4$ in the case II scattering double-magnon states in the XXX chain (symbols) and the corresponding analytical results in the scaling limit (lines).
  The analytical results are $S_{A,13}^{(n),\bos}$ (left), i.e.\ (\ref{BosonSAk1k2n}) and (\ref{BosonSAk1k2}) with $(k_1,k_2)=(1,3)$,
  $S_{A,12}^{(n),\fer}$ (middle), i.e.\ (\ref{FermionSAk1k2n}) and (\ref{FermionSAk1k2}) with $(k_1,k_2)=(1,2)$,
  and $2S_{A,1}^{(n)}$ (right) with $S_{A,1}^{(n)}=-\f1{n-1}\log[x^n+(1-x)^n]$ and $S_{A,1}=-x\log x-(1-x)\log(1-x)$.
  For the exact results in the XXX chain, we have set $L=128$.}
  \label{FigureXXXSAI1I2n}
\end{figure}

The exact results of the R\'enyi and entanglement entropies in the case II scattering double-magnon states of the XXX chain could be calculated either numerically from the local mode method (\ref{SAcInXXX}) and (\ref{SAcIXXX}) or from the analytical expressions (\ref{SAI1I2nXXX}) and (\ref{SAI1I2XXX}).
We compare the results with the corresponding analytical ones in the scaling limit in figure~\ref{FigureXXXSAI1I2n}.

\subsection{General magnon state}

We consider the most general scattering magnon state $|\cI\rag$ (\ref{statepsi}) with Bethe numbers $\cI=\{I_1,I_2,\cdots,I_R\}$, and we focus on the states with finite numbers of magnons. More precisely, we require
\be
\lim_{L\to+\inf} \f{R}{L}=0.
\ee
In the scaling limit, there are fixed scaled Bethe numbers
\be
\io_i = \lim_{L\to+\inf} \f{I_i}{L}, ~~ i=1,2,\cdots,R.
\ee
As there are finite number of magnons, the shift angles approach to the same values as those in the double-magnon state
\be
\lim_{L\to+\inf} \th_{i_1i_2} \to f(\io_{i_1},\io_{i_2}),
\ee
with the function $f(\io_{i_1},\io_{i_2})$ being the same as (\ref{fio1io2}), i.e.\ that in the scaling limit the shift angles between two magnons only depend on the two relevant Bethe numbers not on the Bethe numbers of other magnons.
For example, for the state with seven magnons $(I_1,I_2,I_3,I_4,I_5,I_6,I_7)=\{1,3,\f{L}{4},\f{L}{4}+2,\f{L}{4}+4,\f{L}{2},\f{L}{2}+2\}$ in the scaling limit, we have the shift angles $\th_{i_1i_2}$, $i_1,i_2=1,2,\cdots,6$ forming a $7\times7$ matrix
\be
\th_{i_1i_2} = \lt(\ba{cc|ccc|cc}
0 & 0 & 0         & 0         & 0         & 0        & 0        \\
0 & 0 & 0         & 0         & 0         & 0        & 0        \\ \hline
0 & 0 & 0         & \pi       & \pi       & \pi-\arctan\f43 & \pi-\arctan\f43 \\
0 & 0 & -\pi      & 0         & \pi       & \pi-\arctan\f43 & \pi-\arctan\f43 \\
0 & 0 & -\pi      & -\pi      & 0         & \pi-\arctan\f43 & \pi-\arctan\f43 \\ \hline
0 & 0 & \arctan\f43-\pi & \arctan\f43-\pi & \arctan\f43-\pi & 0        & \pi      \\
0 & 0 & \arctan\f43-\pi & \arctan\f43-\pi & \arctan\f43-\pi & -\pi     & 0
\ea\rt).
\ee
Note that the $7\times7$ matrix could be written into $3\times3$ blocks according to the values of scaled Bethe numbers $\io_i$.

We group the magnons $\cI$ into $\a$ clusters $\cI_a=\{I_{ab}|b=1,2,\cdots,\b_a\}$, $a=1,2,\cdots,\a$, according to the values of the scaled Bethe numbers $\io_a=\lim_{L\to+\inf}\f{I_{ab}}{L}$.
The cluster $a$ has $\b_a$ magnons, and there is $\sum_{a=1}^\a\b_a=R$.
Note that $I\cong I+L$ and $\io\cong \io+1$. The magnons with $\io=0$ and the magnons with $\io=1$ are grouped in the same cluster. The magnons of different clusters have large momentum differences and so decoherent from each other, and in the scaling limit their contributions to the R\'enyi entropy are independent
\be \label{keyresult1}
S_{A,\cI}^{(n),\XXX} = \sum_{a=1}^\a S_{A,\cI_a}^{(n),\XXX}.
\ee
For the cluster of magnons with $\io=0$ and $\io=1$, the R\'enyi entropy is just that in the extremely gapped bosonic chain
\be \label{keyresult2}
S_{A,\cI_a}^{(n),\XXX} = S_{A,K_a}^{(n),\bos}, ~~ K_a=\cI_a.
\ee
For other clusters, the R\'enyi entropy is just that in the extremely gapped fermionic chain
\be \label{keyresult3}
S_{A,\cI_a}^{(n),\XXX} = S_{A,K_a}^{(n),\fer}.
\ee
The momenta $K_a$ are determined by the Bethe numbers $\cI_a=\{I_{ab}|b=1,2,\cdots,\b_a\}$ from (\ref{XXXkifromIi}).
We sort the cluster $a$ Bethe numbers $I_{ab}$ in order from the smallest to the largest, and we get $K_a=\{K_{ab}|b=1,2,\cdots,\b_a\}$ with
\be
K_{ab} = I_{ab}+\f{\b_a+1}{2}-b.
\ee
Especially, for the case with $I_{ab}=I_{a1}+2(b-1)$, there is $K_{ab}=K_{a1}+b-1$.
The formulas (\ref{keyresult1}), (\ref{keyresult2}) and (\ref{keyresult3}) are the key result of the paper.
For example, in the state $|\cI\rag$ with $\cI=\{1,3,\f{L}{4},\f{L}{4}+2,\f{L}{4}+4,\f{L}{2},\f{L}{2}+2\}$, there are the R\'enyi and entanglement entropies in the scaling limit
\be
S_{A,\cI}^{(n),\XXX} = S_{A,13}^{(n),\bos} + S_{A,123}^{(n),\fer} + S_{A,12}^{(n),\fer}.
\ee

Like that in the fermionic and bosonic chains, we further conjecture that in the scaling limit there is
\be \label{SAcI1cupcI2nXXX}
S_{A,\cI\cup\cI'}^{(n),\XXX} = S_{A,\cI}^{(n),\XXX} + S_{A,\cI'}^{(n),\XXX},
\ee
as long as
\be \label{absI1mI2topinf}
| I - I' | \to + \inf, \textrm{~for~all~} I \in \cI \textrm{~and~all~} I' \in \cI'.
\ee
Note that there is $S_{A,G}^{(n),\XXX}=0$ in the ferromagnetic phase of the XXX chain.
The conjecture (\ref{SAcI1cupcI2nXXX}) is more general than (\ref{keyresult1}).
The result (\ref{keyresult1}) is for the case with finite number of magnons, while we conjecture that (\ref{SAcI1cupcI2nXXX}) is valid even for the case with infinite number of magnons in the scaling limit.

\section{XXZ chain}\label{SectionXXZ}

In this section we show that the conclusions derived in the spin-1/2 XXX chain the previous section can be extended to more general interacting quantum Bethe integrable chains such as the spin-1/2 XXZ chain.
The spin-1/2 XXZ chain in transverse field has the Hamiltonian
\be
H = -\f14 \sum_{j=1}^L ( \s_j^x\s_{j+1}^x + \s_j^y\s_{j+1}^y + \D \s_j^z\s_{j+1}^z ) - \f{h}{2} \sum_{j=1}^L \s_j^z.
\ee
We only consider the ferromagnetic phase with $h > \max( 0,\f{1-\D}2)$, and so that the unique ground state is (\ref{XXXG}) and low energy excitations are magnons.

The XXZ chain can be solved with the Bethe ansatz \cite{Gaudin:1983kpuCaux:2014uuq}.
The most general magnon excited state $|\cI\rag$ is still in the form (\ref{statepsi}) with (\ref{statepsicN}) and (\ref{statepsicUj1cdotsjR}).
We require that the state $|\cI\rag$ has finite number of magnons.
The shift angles are determined by the new equation
\be
\ep^{\ii\th_{ii'}} = - \f{1+\ep^{\ii(p_i+p_{i'})}-2\D\ep^{\ii p_{i}}}
                         {1+\ep^{\ii(p_{i}+p_{i'})}-2\D\ep^{\ii p_{i'}}}.
\ee
In the scaling limit $L\to+\inf$, the shift angle between two magnons with Bethe numbers $I_1<I_2$ is
\be
f(\io_1,\io_2) = \f1\ii \log \Big(
                  - \f{1+\ep^{2\pi\ii(\io_1+\io_2)}- 2 \D \ep^{2\pi\ii \io_1}}
                      {1+\ep^{2\pi\ii(\io_1+\io_2)}- 2 \D \ep^{2\pi\ii \io_2}} \Big),
\ee
where the scaled Bethe numbers $\io_1\leq\io_2$ are defined as (\ref{ratiosi1i2}).

\begin{figure}[t]
  \centering
  \includegraphics[height=0.31\textwidth]{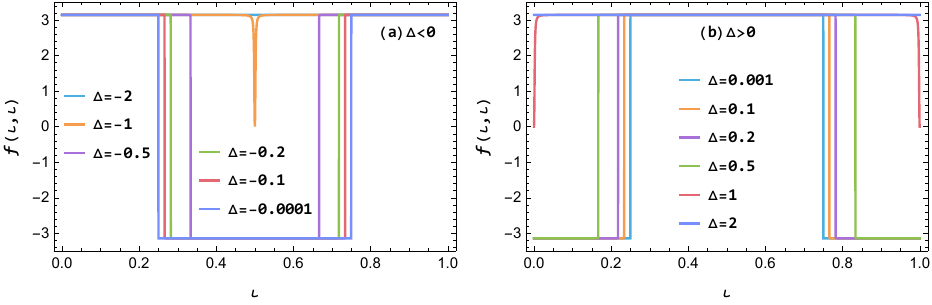}\\
  \caption{The shift angle $\th=f(\io,\io)$ with equal scaled Bethe numbers $\io_1=\io_2=\io$ in the scaling limit of the XXZ chain. In the left panel there is $\D<0$, and in the right panel there is $\D>0$. There is $f(\io,\io)=\pm\pi$ for $\D \neq \pm 1$ in most of the cases. For $\D=-1$ there is $f(\f12,\f12)=0$, for $\D=1$ there is $f(0,0)=f(1,1)=0$.}
  \label{FigureXXZtheta}
\end{figure}

For the R\'enyi and entanglement entropies in the scaling limit, the only relevant shift angles are the ones with equal scaled Bethe numbers $\io_1=\io_2=\io$
\be
f(\io,\io)= \lim_{\e\to0^+} f(\io-\e,\io+\e).
\ee
We plot the shift angle $\th=f(\io,\io)$ with equal scaled Bethe numbers $\io_1=\io_2=\io$ in the scaling limit of the XXZ chain in figure~\ref{FigureXXZtheta}.
For $\D=1$, the XXZ chain becomes the XXX chain, and we have discussed the case in details in the previous section.
For $\D=-1$, there is $f(\io,\io)=\pi$ if $\io\in[0,\f12)\cup(\f12,1]$ and the two magnons behave like fermions, and there is $f(\f12,\f12)=0$ and the two magnons behave like bosons.
For $\D\neq\pm1$, there is always $f(\io,\io)=\pm\pi$, and the two magnons behave like fermions.

The same as that in the XXX chain, we group the magnons in the state $|\cI\rag$ with Bethe numbers $\cI=\{I_1,I_2,\cdots,I_R\}$ into $\a$ clusters $\cI_a$, $a=1,2,\cdots,\a$, according to the values of the scaled Bethe numbers.
As before, the magnons with $\io=0$ and the magnons with $\io=1$ are grouped in the same cluster.
In the scaling limit, the contributions of different clusters to the R\'enyi and entanglement entropies are independent
\be
S_{A,\cI}^{(n),\XXZ} = \sum_{a=1}^\a S_{A,\cI_a}^{(n),\XXZ}.
\ee
For a cluster with scaled Bethe number $\io$, we call it bosonic cluster if $f(\io,\io)=0$, and we call it fermionic cluster if $f(\io,\io)=\pm\pi$.
In figure~\ref{FigureXXZtheta}, we see that in the ferromagnetic phase of the XXZ chain the bosonic cluster only appears at $\D=-1$, $\io=\f12$ and $\D=1$, $\io=0\cong1$.
The contributions of a bosonic cluster is equal to
\be \label{SAcIanXXZbos}
S_{A,\cI_a}^{(n),\XXZ} = S_{A,K_a}^{(n),\bos}, ~~ K_a=\cI_a.
\ee
And the contributions of a fermionic cluster is equal to
\be \label{SAcIanXXZfer}
S_{A,\cI_a}^{(n),\XXZ} = S_{A,K_a}^{(n),\fer},
\ee
where the momenta $K_a$ are determined by (\ref{XXXkifromIi}).

\begin{figure}[t]
  \centering
  \includegraphics[height=0.64\textwidth]{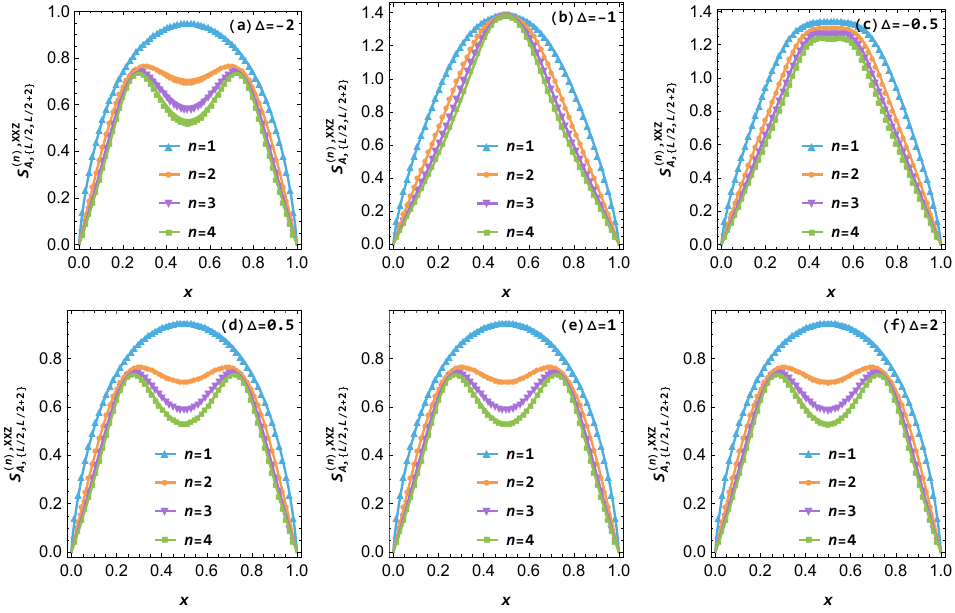}\\
  \caption{The numerical results of R\'enyi and entanglement entropies $S_{A,\{L/2,L/2+2\}}^{(n),\XXZ}$ from the subsystem mode method in the XXZ chain with different values of $\D$ (symbols) and the corresponding analytical results (lines).
  For $\D=-2,0.5,1,2$, the analytical results are $S_{A,12}^{(n),\fer}$, i.e.\ (\ref{FermionSAk1k2n}) and (\ref{FermionSAk1k2}) with $(k_1,k_2)=(1,2)$.
  For $\D=-1$, the analytical result is $S_{A,13}^{(n),\bos}$, i.e.\ (\ref{BosonSAk1k2n}) and (\ref{BosonSAk1k2}) with $(k_1,k_2)=(1,3)$.
  For $\D=-0.5$, the analytical result is $S_{A,14}^{(n),\fer}$, i.e.\ (\ref{FermionSAk1k2n}) and (\ref{FermionSAk1k2}) with $(k_1,k_2)=(1,4)$.
  We have set $L=64$ for the numerical results and $L=+\inf$ for the analytical results.}
  \label{FigureXXZSAI1I2n}
\end{figure}

In figure~\ref{FigureXXZSAI1I2n}, we show the R\'enyi and entanglement entropies $S_{A,\{L/2,L/2+2\}}^{(n),\XXZ}$ in the XXZ chain.
The R\'enyi and entanglement entropies with different values of $\D$ match different analytical expressions according to (\ref{SAcIanXXZbos}) and (\ref{SAcIanXXZfer}).
Especially, when the cluster $\cI=\{L/2,L/2+2\}$ is fermionic, the momenta $K=\{k_1,k_2\}$ is determined by (\ref{XXXkifromIi}), i.e.\ (\ref{k1k2fromI1I2theta}) for two magnons.

The same as that in the XXX chain, we conjecture that in the scaling limit there is
\be
S_{A,\cI\cup\cI'}^{(n),\XXZ} = S_{A,\cI}^{(n),\XXZ} + S_{A,\cI'}^{(n),\XXZ},
\ee
as long as (\ref{absI1mI2topinf}) is satisfied.

\section{Conclusion and discussion} \label{SectionConclusion}

In this paper we proposed three conjectures for the quasiparticle excited state R\'enyi and entanglement entropies of the subsystem $A=[1,\ell]$ in the scaling limit of the general translational invariant quadratic fermionic and bosonic chains of $L$ sites.
The first conjecture is
\bea
&& S_{A,K}^{(n)} - S_{A,G}^{(n)} = - \f{1}{n-1}\log\tr\cR_{A,K}^n, \label{conjecture11}\\
&& S_{A,K} - S_{A,G} = - \tr(\cR_{A,K}\log\cR_{A,K}), \label{conjecture12}
\eea
with the condition
\be \label{condition1}
\max_{k\in K}\f{1}{\ve_k} \ll \min(\ell,L-\ell).
\ee
The second conjecture is
\bea
&& S_{A,K\cup K'}^{(n)} - S_{A,K'}^{(n)} = - \f{1}{n-1}\log\tr\cR_{A,K}^n, \label{conjecture21}\\
&& S_{A,K\cup K'} - S_{A,K'} = - \tr(\cR_{A,K}\log\cR_{A,K}), \label{conjecture22}
\eea
with the condition
\bea
&& \max_{k\in K}\f{1}{\ve_k} \ll \min(\ell,L-\ell), \label{condition21} \\
&& |k-k'|\to+\inf,{\rm~for~all~} k \in K{\rm~and~all~} k' \in K'. \label{condition22}
\eea
The RHS (\ref{conjecture11}), (\ref{conjecture12}), (\ref{conjecture21}) and (\ref{conjecture22}) are calculated in the non-interacting model with the number operator as the Hamiltonian, and we conjecture that these equations are valid as long as the corresponding condition (\ref{condition1}) or [(\ref{condition21}) plus (\ref{condition22})] is satisfied.
The third conjecture is
\bea
&& S_{A,K\cup K'}^{(n)} - S_{A,G}^{(n)} = S_{A,K}^{(n)} + S_{A,K'}^{(n)} - 2 S_{A,G}^{(n)}, \label{conjecture31} \\
&& S_{A,K\cup K'} - S_{A,G} = S_{A,K} + S_{A,K'} - 2 S_{A,G}, \label{conjecture32}
\eea
with the condition
\be \label{condition3}
|k-k'|\to+\inf,{\rm~for~all~} k \in K{\rm~and~all~} k' \in K'.
\ee
The second conjecture could be derived from the first conjecture plus the third conjecture.

We also investigated the magnon excited state R\'enyi and entanglement entropies in the scaling limit of the XXX and XXZ chains. In particular, we showed that as far as the number of excited magnons is small one can decompose the R\'enyi and entanglement entropies to the sum of the R\'enyi and entanglement entropies of particular excited states in the fermionic and bosonic chains with the number operator as the Hamiltonian. It would be interesting to see how this picture changes when one looks to the spinon excited states in which the number of excited modes are proportional to the size of the system. The techniques used in the papers \cite{Miwa:2018cqc,Smirnov:2018oek} might be useful in this direction.

In previous works \cite{Zhang:2020vtc,Zhang:2020dtd} and the present paper, we have focused on one-dimensional quantum systems. The entanglement in excited states of higher dimensional free and interacting quantum systems has been investigated in for example \cite{Castro-Alvaredo:2019lmj,Angel-Ramelli:2020xvd}, suggesting that there are simple formulas as in one dimensions.
It would be interesting to adapt the subsystem mode method to higher dimensions and check the results in various higher dimensional quantum systems.

\section*{Acknowledgements}

We thank Song Cheng for helpful discussions.
We thank Olalla Castro-Alvaredo, Benjamin Doyon, Marcos Rigol and Lev Vidmar for reading a previous version of the draft and helpful comments and suggestions.
MAR thanks ICTP for its hospitality during the visit which part of this work is done.
MAR thanks CNPq and FAPERJ (grant number 210.354/2018) for partial support.

\appendix

\section{Calculations for states in nonorthonormal basis} \label{appNOB}

In this appendix, we give an efficient procedure to calculate the R\'enyi and von Neumann entropies for density matrices in a general nonorthonormal basis.

We consider the general density matrix
\be \label{rcP}
\r_\cP=\sum_{i,j} \cP_{ij} |\phi_i\rag\lag\phi_j|,
\ee
in a general nonorthonormal basis $|\phi_i\rag$ with the positive matrix
\be
\cQ_{ij}=\lag\phi_i|\phi_j\rag.
\ee
It is convenient to define another matrix
\be
\cR=\cP\cQ.
\ee
Here $\r_\cP$ could be the density matrix of the total system or the RDM of a subsystem.
The R\'enyi entropy and von Neumann entropy could be easily written as
\bea
&& S_\cP^{(n)} = - \f{1}{n-1}\log\tr\cR^n, \\
&& S_\cP = - \tr(\cR\log\cR).
\eea

The positive matrix $\cQ$ could be written in terms of the unitary matrix $U$ and the positive diagonal matrix $\Lam$
\be
\cQ = U \Lam U^\dag,
\ee
with $U$ and $\Lam$ being constructed respectively by the eigenvectors and eigenvalues of $\cQ$.
Then the nonorthonormal basis $|\phi_i\rag$ could be written as
\be
|\phi_i\rag = \sum_j (U^* \Lam^{1/2})_{ij} |\psi_j\rag,
\ee
with the orthonormal basis $|\psi_i\rag$ satisfying $\lag\psi_i|\psi_j\rag=\d_{ij}$.
We further write the density matrix as
\be
\r_\cP=\sum_{i,j} \cS_{ij} |\psi_i\rag\lag\psi_j|,
\ee
with
\be
\cS = \Lam^{1/2} U^\dag \cP U \Lam^{1/2}.
\ee
The R\'enyi entropy and von Neumann entropy could be easily written as
\bea
&& S_\cP^{(n)} = - \f{1}{n-1}\log\tr\cS^n, \\
&& S_\cP = - \tr(\cS\log\cS).
\eea

When the matrices $\cP$, $\cQ$, $\cR$, $\cS$ are block diagonal with
\be
\cP = \bigoplus_b \cP_{(b)}, ~~
\cQ = \bigoplus_b \cQ_{(b)}, ~~
\cR = \bigoplus_b \cR_{(b)}, ~~
\cS = \bigoplus_b \cS_{(b)}.
\ee
one could further write the R\'enyi and von Neumann entropies as
\bea
&& S_\cP^{(n)} = - \f{1}{n-1}\log \Big( \sum_b \tr\cR^n_{(b)} \Big) = - \f{1}{n-1}\log \Big( \sum_b \tr\cS^n_{(b)} \Big), \\
&& S_\cP = - \sum_b \tr(\cR_{(b)}\log\cR_{(b)}) = - \sum_b \tr(\cS_{(b)}\log\cS_{(b)}),
\eea

In summary, for every density matrix with matrix $\cP$ in nonorthonormal basis with matrix $\cQ$, we define the matrices $\cR$, $\cS$.
We could calculate the R\'enyi and von Neumann entropies using either $\cR$ or $\cS$.
Actually, anything defined in terms of the density matrices could be calculated from $\cS$.

\section{Free fermions: RDM and entanglement} \label{appFer}

In this appendix we will review and refine the subsystem mode method for the fermionic chain with the number operator (\ref{fermionicchainN}) as the Hamiltonian, or equivalently the extremely gapped limit $\lam\to+\inf$ of the Hamiltonian (\ref{fermionicchainwithlamgam}), and then presents some new forms of the correlation matrix method and finally present some examples.

\subsection{Subsystem mode method}\label{appSubsystem}

The subsystem mode method was used in \cite{Zhang:2020vtc,Zhang:2020dtd} to calculate analytically the R\'enyi entropy with integer index $n\geq2$ in the quasiparticle excited states of the extremely gapped fermionic and bosonic chain.
In this subsection, we will refine the subsystem mode method and show how to use the method to calculate the R\'enyi and entanglement entropies in the quasiparticle excited states of the extremely gapped fermionic chain.
The method is efficient for both analytical and numerical calculations.
In fact, the procedure we prescribe is applicable to any quantity that is defined for the RDMs.

There are subsystem $A=[1,\ell]$ and its complement is $B=[\ell+1,L]$.
We will mainly focus on the results in the scaling limit $L\to+\inf$, $\ell\to+\inf$ with fixed ratio $x=\f{\ell}{L}$.
With the number operator (\ref{fermionicchainN}) as the Hamiltonian, the ground state $|G\rag$ is annihilated by all the local lowering operators
\be
a_j|G\rag=0, ~~ j=1,2,\cdots,L,
\ee
and so it could be written as a direct product form $|G\rag=|G_A\rag\otimes|G_B\rag$ with $a_j|G_A\rag=0$ for all $j\in A$ and $a_j|G_B\rag=0$ for all $j\in B$.
We divide the quasiparticle modes as $c_k=c_{A,k}+c_{B,k}$, $c_k^\dag=c_{A,k}^\dag+c_{B,k}^\dag$ with the subsystem modes
\bea \label{FermionSubsystemModes}
&& c_{A,k} = \f{1}{\sr{L}}\sum_{j\in A} \ep^{-\ii j p_k}a_j, ~~
   c_{A,k}^\dag = \f{1}{\sr{L}}\sum_{j\in A} \ep^{\ii j p_k}a_j^\dag, \nn\\
&& c_{B,k} = \f{1}{\sr{L}}\sum_{j\in B} \ep^{-\ii j p_k}a_j, ~~
   c_{B,k}^\dag = \f{1}{\sr{L}}\sum_{j\in B} \ep^{\ii j p_k}a_j^\dag.
\eea
There are anti-commutation relations
\be
\{ c_{A,k_1}, c_{A,k_2}^\dag \} = \a_{k_1-k_2}, ~~
\{ c_{B,k_1}, c_{B,k_2}^\dag \} = \b_{k_1-k_2},
\ee
with
\bea
&& \a_k = \f{1}{L} \sum_{j=1}^\ell \ep^{-\f{2\pi\ii j k}{L}}
=\lt\{
\ba{cc}
\f{\ell}{L} & k=0\\
\ep^{-\f{\pi\ii k(\ell+1)}{L}} \f{\sin\f{\pi k\ell}{L}}{L\sin\f{\pi k}{L}} & k\neq0
\ea
\rt.\!\!\!, \label{alphak} \\
&& \b_k = \f{1}{L} \sum_{j=\ell+1}^L \ep^{-\f{2\pi\ii j k}{L}}
=\lt\{
\ba{cc}
1-\f{\ell}{L} & k=0\\
\ep^{-\f{\pi\ii k(L+\ell+1)}{L}} \f{\sin\f{\pi k(L-\ell)}{L}}{L\sin\f{\pi k}{L}} & k\neq0
\ea
\rt.\!\!\!. \label{betak}
\eea
For real $k$ there is $\bar\a_k=\a_{-k}$ and $\bar\b_k=\b_{-k}$.
There is $\b_0=1-\a_0$. For $k\in\rZ$ and $k\neq0$ there is $\b_k=-\a_k$.
In the scaling limit $L\to+\inf$, $\ell\to+\inf$ with fixed $x=\f{\ell}{L}$ and fixed $k$, we have
\bea
&& \a_k
=\lt\{
\ba{cc}
x & k=0\\
\ep^{-\pi\ii k x} \f{\sin(\pi k x)}{\pi k} & k\neq0
\ea
\rt.\!\!\!,\\
&& \b_k
=\lt\{
\ba{cc}
1-x & k=0\\
\ep^{-\pi\ii k (1+x)} \f{\sin[\pi k (1+x)]}{\pi k} & k\neq0
\ea
\rt.\!\!\!.
\eea

We consider an ordered set of momenta $K=\{k_1,\cdots,k_r\}$ with $k_1<\cdots<k_r$.
There is the number of quasiparticles $|K|=r$.
We define the products of subsystem modes
\bea
&& c^\dag_{A,K} = c^\dag_{A,k_1}\cdots c^\dag_{A,k_r}, ~~
   c_{A,K} = (c^\dag_{A,K})^\dag = c_{A,k_r} \cdots c_{A,k_1}, \nn\\
&& c^\dag_{B,K} = c^\dag_{B,k_1}\cdots c^\dag_{B,k_r}, ~~
   c_{B,K} = (c^\dag_{B,K})^\dag = c_{B,k_r} \cdots c_{B,k_1}.
\eea
We emphasize that all the sets used in this subsection are ordered.
The excited state (\ref{fermionKrag}) is
\be
|K\rag=\sum_{K' \subseteq K} s_{K,K'} c^\dag_{A,K'} c^\dag_{B,K \bs K'}|G\rag,
\ee
with $K \bs K'$ being the complement set of $K'$ contained in $K$.
We have defined the factor which is vanishing or just a sign
\be
s_{K,K'} = \lt\{
\ba{cl}
0                             & K' \nsubseteq K \\
{\rm sig}[K',K\bs K']   & K' \subseteq K
\ea
\rt.\!\!\!,
\ee
with ${\rm sig}[K',K\bs K']$ denoting the signature of the two ordered sets $K'$ and $K\bs K'$ joining together without changing the orders of the momenta in each of them.
For example, there are $s_{\{1,2,3\},\{1\}}=s_{\{1,2,3\},\{2,3\}}=1$, $s_{\{1,2,3\},\{2\}}=s_{\{1,2,3\},\{1,3\}}=-1$.
Then we get the RDM
\be
\r_{A,K}=
\sum_{K_1,K_2\subseteq K}
s_{K,K_1} s_{K,K_2}
\lag c_{B,K\bs K_2} c^\dag_{B,K\bs K_1} \rag_G
c^\dag_{A,K_1}
|G_A\rag\lag G_A|
c_{A,K_2}.
\ee

In the nonorthonormal basis $c^\dag_{A,K'}|G_A\rag$ with $K'\subseteq K$ we write the RDM in the form of (\ref{rcP}).
Note that the set $K$ has $2^{|K|}$ different subsets.
We have the RDM
\be \label{rhoAK}
\r_{A,K}=
\sum_{K_1,K_2\subseteq K}
[\cP_{A,K}]_{K_1K_2}
c^\dag_{A,K_1}
|G_A\rag\lag G_A|
c_{A,K_2},
\ee
with the entries of the $2^{|K|}\times2^{|K|}$ matrix $\cP_{A,K}$
\be
[\cP_{A,K}]_{K_1K_2} = s_{K,K_1} s_{K,K_2} \lag c_{B,K\bs K_2} c^\dag_{B,K\bs K_1} \rag_G,
\ee
There is also the $2^{|K|}\times2^{|K|}$ matrix $\cQ_{A,K}$ with entries
\be
[\cQ_{A,K}]_{K_1K_2} = \lag c_{A,K_1} c^\dag_{A,K_2} \rag_G.
\ee


We need to evaluate the expectation values $\lag c_{A,K_1} c^\dag_{A,K_2} \rag_G$ and $\lag c_{B,K_1} c^\dag_{B,K_2} \rag_G$ in the ground state, which are just the determinants
\bea
&& \lag c_{A,K_1} c^\dag_{A,K_2} \rag_G =
\lt\{
\ba{cl}
0                  & |K_1| \neq |K_2| \\
\det \cA_{K_1 K_2} & |K_1| = |K_2|
\ea
\rt.\!\!\!, \nn\\
&& \lag c_{B,K_1} c^\dag_{B,K_2} \rag_G =
\lt\{
\ba{cl}
0                  & |K_1| \neq |K_2| \\
\det \cB_{K_1 K_2} & |K_1| = |K_2|
\ea
\rt.\!\!\!.
\eea
where the $|K_1|\times|K_2|$ matrices $\cA_{K_1 K_2}$ and $\cB_{K_1 K_2}$ have the entries
\be \label{cAK1K2cBK1K2definition}
[\cA_{K_1 K_2}]_{i_1i_2} = \a_{k_{i_1}-k_{i_2}}, ~~
[\cB_{K_1 K_2}]_{i_1i_2} = \b_{k_{i_1}-k_{i_2}},
\ee
with $i_1=1,2,\cdots,|K_1|$, $i_2=1,2,\cdots,|K_2|$, $k_{i_1}\in K_1$, $k_{i_2}\in K_2$ and the definitions of $\a_k$ and $\b_k$ in (\ref{alphak}) and (\ref{betak}).

With the matrices $\cP_{A,K}$ and $\cQ_{A,K}$, we follow the procedure in appendix~\ref{appNOB}, calculate $\cR_{A,K}=\cP_{A,K}\cQ_{A,K}$ and obtain the R\'enyi and entanglement entropies
\bea
&& S_{A,K}^{(n),\fer} = - \f{1}{n-1}\log\tr\cR_{A,K}^n, \\
&& S_{A,K}^\fer = - \tr(\cR_{A,K}\log\cR_{A,K}).
\eea
Note that the matrices $\cP_{A,K}$ and $\cQ_{A,K}$ are block diagonal.
The method is efficient for both analytical and numerical calculations.
When only a few quasiparticles are excited, we could calculate the results analytically, and when more quasiparticles are excited we have to calculate the results numerically.


\subsection{Correlation matrix method}\label{appCorrelation}

To verify the results from the subsystem mode method, we calculate numerically the results using the correlation matrix method \cite{Chung:2001oyk,Vidal:2002rm,Peschel:2002jhw,Latorre:2003kg,Alba:2009th,Alcaraz:2011tn,Berganza:2011mh}.
In the generally gapped fermionic chain, one could define the Majorana modes
\be
d_{2j-1} = a_j + a_j^\dag, ~~
d_{2j} = \ii ( a_j - a_j^\dag ).
\ee
In the general excited state $|K\rag=|k_1k_2\cdots k_r\rag$, one defines the $2\ell \times 2\ell$ correlation matrix $\G^K$ as
\be
\G^K_{m_1m_2} = \lag d_{m_1} d_{m_2} \rag_K - \d_{m_1m_2}, ~~
m_1,m_2 = 1,2,\cdots,2\ell.
\ee
The R\'enyi and entanglement entropies in state $|K\rag$ are
\bea
&& S_{A,K}^{(n),\fer}
   = - \f{1}{2(n-1)} \log \det \Big[ \Big( \f{1+\G^K}{2} \Big)^n + \Big( \f{1 - \G^K}{2} \Big)^n \Big] \nn\\
&& \phantom{S_{A,K}^{(n),\fer}}
   = - \f{1}{2(n-1)} \tr \log \Big[ \Big( \f{1+\G^K}{2} \Big)^n + \Big( \f{1 - \G^K}{2} \Big)^n \Big], \\
&& S_{A,K}^\fer = - \tr\Big( \f{1+\G^K}{2} \log \f{1+\G^K}{2} \Big)
           = - \tr\Big( \f{1-\G^K}{2} \log \f{1-\G^K}{2} \Big).
\eea

With the number operator (\ref{fermionicchainN}) as the Hamiltonian, one defines the $\ell \times\ell$ correlation matrix $C^K$ as
\be
C^K_{j_1j_2} = \lag a_{j_1}^\dag a_{j_2} \rag_K = h^K_{j_2-j_1}, ~~ j_1,j_2 = 1,2,\cdots,\ell,
\ee
with the function
\be
h^K_j = \f{1}{L} \sum_{k\in K} \ep^{\ii j p_k}.
\ee
The R\'enyi and entanglement entropies in state $|K\rag$ are
\bea
&& S_{A,K}^{(n),\fer}
   = - \f{1}{n-1} \log \det [ ( C^K )^n + ( 1 - C^K )^n ] \nn\\
&& \phantom{S_{A,K}^{(n),\fer}}
   = - \f{1}{n-1} \tr \log [ ( C^K )^n + ( 1 - C^K )^n ], \\
&& S_{A,K}^\fer = \tr[ - C^K \log C^K - (1-C^K) \log (1-C^K) ].
\eea

Using the position-momentum duality \cite{Lee:2014nra,Carrasco:2017eul}, we could also write the R\'enyi and entanglement entropies as
\bea
&& S_{A,K}^{(n),\fer}
   = - \f{1}{n-1} \log \det [ ( \cA_\cK )^n + ( 1 - \cA_\cK )^n ] \nn\\
&& \phantom{S_{A,K}^{(n),\fer}}
   = - \f{1}{n-1} \tr \log [ ( \cA_\cK )^n + ( 1 - \cA_\cK )^n ], \label{FermionRenyi}\\
&& S_{A,K}^\fer = \tr[ - \cA_\cK \log \cA_\cK - (1-\cA_\cK) \log (1-\cA_\cK) ], \label{FermionvonNeumann}
\eea
where the $|K|\times|K|$ matrix $\cA_K$ has entries
\be \label{cAKi1i2definition}
[\cA_K]_{{i_1}{i_2}} = \a_{k_{i_1} - k_{i_2}},
\ee
with $i_1,i_2=1,2,\cdots,|K|$, $k_{i_1},k_{i_2} \in K$ and the definition of $\a_k$ in (\ref{alphak}).
We may also define the $|K|\times|K|$ matrix $\cB_K$ which has entries
\be \label{cBKi1i2definition}
[\cB_K]_{{i_1}{i_2}} = \b_{k_{i_1} - k_{i_2}},
\ee
with $i_1,i_2=1,2,\cdots,|K|$, $k_{i_1},k_{i_2} \in K$ and the definition of $\b_k$ in (\ref{betak}).
In fact, there are $\cA_K=\cA_{KK}$, $\cB_K=\cB_{KK}$ with the definitions of $\cA_{K_1K_2}$, $\cB_{K_1K_2}$ in (\ref{cAK1K2cBK1K2definition}).
As there is always $k_{i_1} - k_{i_2} \in \rZ$ in the fermionic chain, we have
\be
\cB_K = 1 - \cA_K.
\ee

Using the determinant of the block matrix
\be
\det \lt( \ba{cc} \cX & \cY \\ \cZ & \cW \ea\rt) = \det \cX \det( \cW - \cZ\cX^{-1}\cY ),
\ee
and the fact that the matrices $\cA_K$ and $\cB_K$ commute, we could write the R\'enyi entropy (\ref{FermionRenyi}) as%
\footnote{The solves the puzzle in footnote 4 of \cite{Zhang:2020dtd}.}
\be \label{FermionRenyiII}
S_{A,K}^{(n),\fer} = - \f{1}{n-1} \log \det \Psi_{A,K}^{(n)},
\ee
with the $n|K|\times n|K|$ matrix $\Psi_{A,K}^{(n)}$ written in $n\times n$ blocks as
\be
\Psi_{A,K}^{(n)} = \lt(\ba{cccc}
\cB_{K}          & \cA_{K} &        &          \\
                 & \cB_{K} & \ddots &         \\
                 &         & \ddots & \cA_{K} \\
(-)^{n-1}\cA_{K} &         &        & \cB_{K} \\
\ea\rt).
\ee

The formula of the R\'enyi entropy (\ref{FermionRenyi}) is more efficient than (\ref{FermionRenyiII}). The formula (\ref{FermionRenyi}) is applicable for any index $n\neq1$ and from the formula one could do analytical continuation $n\to1$ and get the entanglement entropy (\ref{FermionvonNeumann}).
On the other hand, the formula (\ref{FermionRenyiII}) is only applicable for integer $n>1$, and it is not apparent how to do analytical continuation from it.

\subsection{Examples of RDM and entanglement entropy in free fermions}\label{appExamples}

In this appendix we give a few examples of the RDM and entanglement entropy for the quasiparticle excited states of the Hamiltonian (\ref{fermionicchainN}).

\subsubsection{Ground state}\label{subsecfermiong}

In the extremely gapped limit, the ground state is a direct product state
\be
|G\rag = |G_A\rag |G_B\rag.
\ee
The R\'enyi and entanglement entropies are trivially vanishing
\be
S_{A,G}^{(n),\fer}=S_{A,G}^\fer=0.
\ee

\subsubsection{Single-particle states}\label{subsecfermionk}

In the single-particle state $|k\rag$, there is the RDM
\be
\r_{A,k} = (1-x) |G_A\rag\lag G_A| + c_{A,k}^\dag |G_A\rag\lag G_A| c_{A,k}.
\ee
In the nonorthonormal basis
\be
|\phi_0\rag = |G_A\rag, ~~
|\phi_1\rag = c^\dag_{A,k}|G_A\rag,
\ee
there are matrices
\be
\cP_{A,k}=\diag(1-x,1), ~~
\cQ_{A,k}=\diag(1,x), ~~
\cR_{A,k}=\diag(1-x,x).
\ee
The R\'enyi and entanglement entropies are
\bea
&& S_{A,k}^{(n),\fer} = -\f{1}{n-1}\log[x^n+(1-x)^n], \label{FermionSAkn}\\
&& S_{A,k}^\fer=-x\log x-(1-x)\log(1-x). \label{FermionSAk}
\eea
These results have been obtained in \cite{Pizorn:2012aut,Berkovits:2013mii,Molter2014Bound,Castro-Alvaredo:2018dja,%
Castro-Alvaredo:2018bij}.

\subsubsection{Double-particle states}

In the single-particle state $|k_1k_2\rag$, there is the RDM
\bea
&& \r_{A,k_1k_2} = [ (1-x)^2 - |\a_{12}|^2 ] |G_A\rag\lag G_A|
              + ( 1-x ) c_{A,k_1}^\dag |G_A\rag\lag G_A| c_{A,k_1} \nn\\
&& \phantom{\r_{A,k_1k_2} =}
              + \a_{12} c_{A,k_1}^\dag |G_A\rag\lag G_A| c_{A,k_2}
              + \bar\a_{12} c_{A,k_2}^\dag |G_A\rag\lag G_A| c_{A,k_1}  \nn\\
&& \phantom{\r_{A,k_1k_2} =}
              + ( 1-x ) c_{A,k_2}^\dag |G_A\rag\lag G_A| c_{A,k_2}
              + c_{A,k_1}^\dag c_{A,k_2}^\dag |G_A\rag\lag G_A| c_{A,k_2} c_{A,k_1},
\eea
with $\a_{12} \equiv \a_{k_1-k_2}$. We have used $\b_k=-\a_{k}$ for nonvanishing integer $k$.
In the nonorthonormal basis
\bea \label{Fermionphi0123}
&& |\phi_0\rag = |G_A\rag, \nn\\
&& |\phi_1\rag = c^\dag_{A,k_1}|G_A\rag, \nn\\
&& |\phi_2\rag = c^\dag_{A,k_2}|G_A\rag, \nn\\
&& |\phi_3\rag = c^\dag_{A,k_1}c^\dag_{A,k_2}|G_A\rag,
\eea
there are matrices
\bea
&& \cP_{A,k_1k_2}=\lt(
\ba{cccc}
(1-x)^2-|\a_{12}|^2 & 0           & 0       & 0 \\
0                   & 1-x         & \a_{12} & 0 \\
0                   & \bar\a_{12} & 1-x     & 0 \\
0                   & 0           &         & 1
\ea
\rt), \nn\\
&& \cQ_{A,k_1k_2}=\lt(
\ba{cccc}
1 & 0           & 0       & 0 \\
0 & x           & \a_{12} & 0 \\
0 & \bar\a_{12} & x       & 0 \\
0 & 0           &         & x^2-|\a_{12}|^2
\ea
\rt),\\
&& \cR_{A,k_1k_2}=\lt(
\ba{cccc}
(1-x)^2-|\a_{12}|^2 & 0                  & 0                  & 0 \\
0                   & x(1-x)+|\a_{12}|^2 & \a_{12}            & 0 \\
0                   & \bar\a_{12}        & x(1-x)+|\a_{12}|^2 & 0 \\
0                   & 0                  &                    & x^2-|\a_{12}|^2
\ea
\rt).\nn
\eea

Alternatively, in the orthonormal basis
\bea \label{Fermionpsi0123}
&& |\psi_0\rag = |G_A\rag, \nn\\
&& |\psi_1\rag = \f{1}{\sr{2|\a_{12}|(x+|\a_{12}|)}} ( \sr{\a_{12}} c_{A,k_1}^\dag + \sr{\bar\a_{12}} c_{A,k_2}^\dag ) |G_A\rag,\nn\\
&& |\psi_2\rag = \f{1}{\sr{2|\a_{12}|(x-|\a_{12}|)}} ( \sr{\a_{12}} c_{A,k_1}^\dag - \sr{\bar\a_{12}} c_{A,k_2}^\dag ) |G_A\rag,\nn\\
&& |\psi_3\rag = \f{1}{\sr{x^2-|\a_{12}|^2}} c_{A,k_1}^\dag c_{A,k_2}^\dag |G_A\rag,
\eea
we have the RDM
\be
\r_{A,k_1k_2} = \diag[
(1-x)^2-|\a_{12}|^2,
(x+|\a_{12}|)(1-x+|\a_{12}|),
(x-|\a_{12}|)(1-x-|\a_{12}|),
x^2-|\a_{12}|^2 ].
\ee

In either the nonorthonormal basis (\ref{Fermionphi0123}) or the orthonormal basis (\ref{Fermionpsi0123}), we get the R\'enyi and entanglement entropies in the double particle state $|k_1k_2\rag$
\bea
&& \hspace{-8mm}
   S_{A,k_1k_2}^{(n),\fer} = - \f{1}{n-1} \log
                  \{ [ (x + |\a_{12}|)^n + (1 - x - |\a_{12}|)^n ]
                     [ (x - |\a_{12}|)^n + (1 - x + |\a_{12}|)^n ] \}, \label{FermionSAk1k2n} \\
&& \hspace{-8mm}
   S_{A,k_1k_2}^\fer = - (x + |\a_{12}|) \log (x + |\a_{12}|)
               - (1 - x - |\a_{12}|) \log (1 - x - |\a_{12}|) \nn\\
&& \hspace{-8mm}\phantom{S_{A,k_1k_2}^\fer =}
               - (x - |\a_{12}|) \log (x - |\a_{12}|)
               - (1 - x + |\a_{12}|) \log (1 - x + |\a_{12}|). \label{FermionSAk1k2}
\eea
It is remarkable that the R\'enyi entropy is written as an explicit analytical function of the general R\'enyi index $n$, and this allows us to get the analytical form of the entanglement entropy.
The R\'enyi and entanglement entropies (\ref{FermionSAk1k2n}) and (\ref{FermionSAk1k2}) have been obtained in \cite{Zhang:2020vtc,Zhang:2020dtd} from formulas (\ref{FermionRenyi}) and (\ref{FermionvonNeumann}).

\section{Free bosons: RDM and entanglement}\label{appBos}

In this appendix we will review and refine the subsystem mode method for the bosonic chain with the number operator (\ref{uncoupled-bosonicchain}) as the Hamiltonian, or equivalently the extremely gapped limit $m\to+\inf$ of the Hamiltonian (\ref{KGbosonhamiltonian}), and then present some new forms of the wave function method and finally present some examples.

\subsection{Subsystem mode method}\label{appSubsystem2}

The analytical calculation using the subsystem mode method is similar to that in the fermionic chain in the previous appendix, and we will keep it brief in this subsection.
We divide the quasiparticle modes $c_k,c^\dag_k$ into the subsystem modes
\bea \label{BosonSubsystemModes}
&& c_{A,k} = \f{1}{\sr{L}}\sum_{j\in A} \ep^{-\ii j p_k} a_j, ~~
   c_{A,k}^\dag = \f{1}{\sr{L}}\sum_{j\in A} \ep^{\ii j p_k} a_j^\dag, \nn\\
&& c_{B,k} = \f{1}{\sr{L}}\sum_{j\in B} \ep^{-\ii j p_k} a_j, ~~
   c_{B,k}^\dag = \f{1}{\sr{L}}\sum_{j\in B} \ep^{\ii j p_k} a_j^\dag.
\eea
There are commutation relations
\be
[ c_{A,k_1}, c_{A,k_2}^\dag ] = \a_{k_1-k_2}, ~~
[ c_{B,k_1}, c_{B,k_2}^\dag ] = \b_{k_1-k_2},
\ee
with $\a_k$ and $\b_k$ defined the same as those in (\ref{alphak}) and (\ref{betak}).

For an arbitrary set $K=\{k_1^{r_1},\cdots,k_s^{r_s}\}$, we have the number of momenta
\be \label{Rdefinition}
R = |K| = \sum_{i=1}^s r_s.
\ee
We define the products of subsystem modes (\ref{BosonSubsystemModes})
\bea
&& c^\dag_{A,K} = (c^\dag_{A,k_1})^{r_1}\cdots (c^\dag_{A,k_s})^{r_s}, ~~
   c_{A,K} = (c^\dag_{A,K})^\dag = (c_{A,k_s})^{r_s} \cdots (c_{A,k_1})^{r_1}, \nn\\
&& c^\dag_{B,K} = (c^\dag_{B,k_1})^{r_1}\cdots (c^\dag_{B,k_s})^{r_s}, ~~
   c_{B,K} = (c^\dag_{B,K})^\dag = (c_{B,k_s})^{r_s} \cdots (c_{B,k_1})^{r_1}.
\eea
Then there is the excited state
\be
|K\rag=\sum_{K' \subseteq K} s_{K,K'} c^\dag_{A,K'} c^\dag_{B,K \bs K'}|G\rag,
\ee
with $K \bs K'$ being the complement of $K'$ in $K$ and the factor $s_{K,K'}$ being defined as
\be
s_{K,K'} = \lt\{
\ba{cl}
0                                             & K' \nsubseteq K \\
\prod_{i=1}^s \f{\sr{r_i!}}{r'_i!(r_i-r'_i)!} & K' \subseteq K
\ea
\rt.\!\!\!,
\ee
Then we get the RDM
\be
\r_{A,K}=
\sum_{K_1,K_2\subseteq K}
s_{K,K_1} s_{K,K_2}
\lag c_{B,K\bs K_2} c^\dag_{B,K\bs K_1} \rag_G
c^\dag_{A,K_1}
|G_A\rag\lag G_A|
c_{A,K_2}.
\ee
Note the possible momenta repetitions of the sets and subsets used in this paper, and one need to be careful to calculate the subsets and their complements.
The general sets $K=\{k_1^{r_1},\cdots,k_s^{r_s}\}$ has $d(K)=\prod_{i=1}^s (r_i+1)$ different subsets.
For example, the set $K=\{1^2,2\}$ has $R=3$ quasiparticles and $d(K)=6$ different subsets. Its subsets $K'$ and their corresponding complements $K \bs K'$ are
\be
\begin{array}{|c|c|c|c|c|c|c|}\hline
  K'      & \varnothing & \{1\}   & \{2\}   & \{1^2\} & \{1,2\} & \{1^2,2\} \\\hline
  K\bs K' & \{1^2,2\}   & \{1,2\} & \{1^2\} & \{2\}   & \{1\}   & \varnothing\\\hline
\end{array}.
\ee

Then we write the RDM in the form of (\ref{rcP}).
We have
\be
\r_{A,K}=
\sum_{K_1,K_2\subseteq K}
[\cP_{A,K}]_{K_1,K_2}
c^\dag_{A,K_1}
|G_A\rag\lag G_A|
c_{A,K_2},
\ee
with the entries of the $d(K)\times d(K)$ matrix $\cP_{A,K}$
\be
[\cP_{A,K}]_{K_1K_2} = s_{K,K_1} s_{K,K_2} \lag c_{B,K\bs K_2} c^\dag_{B,K\bs K_1} \rag_G,
\ee
There is also the $d(K)\times d(K)$ matrix $\cQ_{A,K}$ with entries
\be
[\cQ_{A,K}]_{K_1K_2} = \lag c_{A,K_1} c^\dag_{A,K_2} \rag.
\ee

We need to evaluate the expectation values $\lag c_{A,K_1} c^\dag_{A,K_2} \rag_G$ and $\lag c_{B,K_1} c^\dag_{B,K_2} \rag_G$, which are just the permanents
\bea
&& \lag c_{A,K_1} c^\dag_{A,K_2} \rag_G =
\lt\{
\ba{cl}
0                  & |K_1| \neq |K_2| \\
\per \cA_{K_1 K_2} & |K_1| = |K_2|
\ea
\rt.\!\!\!, \nn\\
&& \lag c_{B,K_1} c^\dag_{B,K_2} \rag_G =
\lt\{
\ba{cl}
0                  & |K_1| \neq |K_2| \\
\per \cB_{K_1 K_2} & |K_1| = |K_2|
\ea
\rt.\!\!\!.
\eea
where  $|K_1|\times|K_2|$ matrices $\cA_{K_1 K_2}$ and $\cB_{K_1 K_2}$ have the same entries as (\ref{cAK1K2cBK1K2definition}).

With the matrices $\cP_{A,K}$ and $\cQ_{A,K}$, we get $\cR_{A,K}=\cP_{A,K}\cQ_{A,K}$ and the R\'enyi and entanglement entropies
\bea
&& S_{A,K}^{(n),\bos} = - \f{1}{n-1}\log\tr\cR_{A,K}^n, \\
&& S_{A,K}^\bos = - \tr(\cR_{A,K}\log\cR_{A,K}).
\eea

\subsection{Wave function method}\label{appwave-function}

We could also calculate the Schatten distance from the wave function method \cite{Castro-Alvaredo:2018dja,Castro-Alvaredo:2018bij}. One could also see \cite{Zhang:2020txb,Zhang:2020dtd}.
With the number operator (\ref{uncoupled-bosonicchain}) as the Hamiltonian, there sets up a permanent formula for the R\'enyi entropy \cite{Zhang:2020dtd}
\be \label{BosonRenyiPermanentFormula}
S_{A,K}^{(n),\bos} = - \f{1}{n-1} \log \f{\per \O_{A,K}^{(n)}}{N_K^n},
\ee
with the $n|K|\times n|K|$ matrix $\Omega_{A,K}^{(n)}$ written in $n\times n$ blocks as
\be
\O_{A,K}^{(n)} = \lt(\ba{cccc}
\cB_{K} & \cA_{K} &        &          \\
        & \cB_{K} & \ddots &         \\
        &         & \ddots & \cA_{K} \\
\cA_{K} &         &        & \cB_{K} \\
\ea\rt),
\ee
and the $|K|\times |K|$ matrices $\cA_{K}$ and $\cB_{K}$ defined the same as (\ref{cAKi1i2definition}) and (\ref{cBKi1i2definition}).
The formula of the R\'enyi entropy (\ref{BosonRenyiPermanentFormula}) is applicable only for integer $n>1$ and it is not apparent how to do analytical continuation $n\to1$ to get the entanglement entropy.

\subsection{Examples of RDM and entanglement entropy in free bosons}\label{appExamples2}

In this appendix we give a few examples of the RDM and entanglement entropy for the quasiparticle excited states of the Hamiltonian (\ref{fermionicchainN}) for free bosons.

\subsubsection{Ground state}

It is the same as that in the extremely gapped fermionic chain in subsection~\ref{subsecfermiong}.
The R\'enyi and entanglement entropies are
\be
S_{A,G}^{(n),\bos}=S_{A,G}^\bos=0.
\ee

\subsubsection{Single-particle states}

It is the same as that in the extremely gapped fermionic chain in subsection~\ref{subsecfermionk}.
The R\'enyi and entanglement entropies are
\bea
&& S_{A,k}^{(n),\bos} = -\f{1}{n-1}\log[x^n+(1-x)^n], \label{BosonSAkn} \\
&& S_{A,k}^\bos=-x\log x-(1-x)\log(1-x). \label{BosonSAk}
\eea

\subsubsection{Double-particle states with equal momenta}

For the double-particle state with the same momenta, there is the RDM
\be
\r_{A,k^2} =    (1-x)^2 |G_A\rag\lag G_A|
            + 2 (1-x) c_{A,k}^\dag |G_A\rag\lag G_A| c_{A,k}
            + \f12 (c_{A,k}^\dag)^2 |G_A\rag\lag G_A| c_{A,k}^2.
\ee
In the nonorthonormal basis
\be
|\phi_0\rag = |G_A\rag, ~~
|\phi_1\rag = c^\dag_{A,k}|G_A\rag, ~~
|\phi_2\rag = \f12 (c_{A,k}^\dag)^2 |G_A\rag,
\ee
we have the matrices
\bea
&& \cP_{A,k^2} = \diag[(1-x)^2,2(1-x),1], ~~\\
&& \cQ_{A,k^2} = \diag(1,x,x^2), ~~\\
&& \cR_{A,k^2} = \diag[(1-x)^2,2x(1-x),x^2].
\eea
We get the R\'enyi and entanglement entropies
\bea
&& S_{A,k^2}^{(n),\bos} = - \f{1}{n-1}\log\{x^{2n}+[2x(1-x)]^n+(1-x)^{2n}\}, \label{BosonSAk2n}\\
&& S_{A,k^2}^\bos = -x^2\log (x^2)-2x(1-x)\log[2x(1-x)]-(1-x)^2\log[(1-x)^2], \label{BosonSAk2}
\eea
which are just the universal R\'enyi and entanglement entropies found in \cite{Castro-Alvaredo:2018dja,Castro-Alvaredo:2018bij}.

\subsubsection{Double-particle states with different momenta}

In bosonic chain with the number operator (\ref{uncoupled-bosonicchain}) as the Hamiltonian, there is the RDM \cite{Zhang:2020dtd}
\bea
&& \r_{A,k_1k_2} = [ (1-x)^2 + |\a_{12}|^2 ] |G_A\rag\lag G_A|
              + ( 1-x ) c_{A,k_1}^\dag |G_A\rag\lag G_A| c_{A,k_1} \nn\\
&& \phantom{\r_{A,k_1k_2} =}
              - \a_{12} c_{A,k_1}^\dag |G_A\rag\lag G_A| c_{A,k_2}
              - \bar\a_{12} c_{A,k_2}^\dag |G_A\rag\lag G_A| c_{A,k_1}  \nn\\
&& \phantom{\r_{A,k_1k_2} =}
              + ( 1-x ) c_{A,k_2}^\dag |G_A\rag\lag G_A| c_{A,k_2}
              + c_{A,k_1}^\dag c_{A,k_2}^\dag |G_A\rag\lag G_A| c_{A,k_2} c_{A,k_1}.
\eea
In the nonorthonormal basis
\bea \label{Bosonphi0123}
&& |\phi_0\rag = |G_A\rag, \nn\\
&& |\phi_1\rag = c^\dag_{A,k_1}|G_A\rag, \nn\\
&& |\phi_2\rag = c^\dag_{A,k_2}|G_A\rag, \nn\\
&& |\phi_3\rag = c^\dag_{A,k_1}c^\dag_{A,k_2}|G_A\rag,
\eea
we have the matrices
\bea
&& \cP_{A,k_1k_2}=\lt(
\ba{cccc}
(1-x)^2+|\a_{12}|^2 & 0            & 0        & 0 \\
0                   & 1-x          & -\a_{12} & 0 \\
0                   & -\bar\a_{12} & 1-x      & 0 \\
0                   & 0            &          & 1
\ea
\rt), \nn\\
&& \cQ_{A,k_1k_2}=\lt(
\ba{cccc}
1 & 0           & 0       & 0 \\
0 & x           & \a_{12} & 0 \\
0 & \bar\a_{12} & x       & 0 \\
0 & 0           &         & x^2+|\a_{12}|^2
\ea
\rt), \nn\\
&& \cR_{A,k_1k_2}=\lt(
\ba{cccc}
(1-x)^2+|\a_{12}|^2 & 0                  & 0                  & 0 \\
0                   & x(1-x)-|\a_{12}|^2 & (1-2x)\a_{12}            & 0 \\
0                   & (1-2x)\bar\a_{12}        & x(1-x)-|\a_{12}|^2 & 0 \\
0                   & 0                  &                    & x^2+|\a_{12}|^2
\ea
\rt).
\eea

Alternatively, in the orthonormal basis
\bea \label{Bosonpsi0123}
&& |\psi_0\rag = |G_A\rag, \nn\\
&& |\psi_1\rag = \f{1}{\sr{2|\a_{12}|(x+|\a_{12}|)}} ( \sr{\a_{12}} c_{A,k_1}^\dag + \sr{\bar\a_{12}} c_{A,k_2}^\dag ) |G_A\rag,\nn\\
&& |\psi_2\rag = \f{1}{\sr{2|\a_{12}|(x-|\a_{12}|)}} ( \sr{\a_{12}} c_{A,k_1}^\dag - \sr{\bar\a_{12}} c_{A,k_2}^\dag ) |G_A\rag,\nn\\
&& |\psi_3\rag = \f{1}{\sr{x^2+|\a_{12}|^2}} c_{A,k_1}^\dag c_{A,k_2}^\dag |G_A\rag,
\eea
we have the RDM
\be
\r_{A,k_1k_2} = \diag[
(1-x)^2+|\a_{12}|^2,
(x+|\a_{12}|)(1-x-|\a_{12}|),
(x-|\a_{12}|)(1-x+|\a_{12}|),
x^2+|\a_{12}|^2 ].
\ee

In either the nonorthonormal basis (\ref{Bosonphi0123}) or the orthonormal basis (\ref{Bosonpsi0123}), we get the R\'enyi entropy with general index $n$
\bea \label{BosonSAk1k2n}
&& S_{A,k_1k_2}^{(n),\bos} =  - \f{1}{n-1} \log
                  \{ ( x^2+|\a_{12}|^2 )^n
                   + [ (1-x)^2+|\a_{12}|^2 ]^n \nn\\
&& \phantom{S_{A,k_1k_2}^{(n),\bos} =}
                   + (x+|\a_{12}|)^n(1-x-|\a_{12}|)^n
                   + (x-|\a_{12}|)^n(1-x+|\a_{12}|)^n \},
\eea
and the entanglement entropy
\bea \label{BosonSAk1k2}
&& S_{A,k_1k_2}^\bos = - ( x^2+|\a_{12}|^2 ) \log ( x^2+|\a_{12}|^2 )
                     - [ (1-x)^2+|\a_{12}|^2 ] \log [ (1-x)^2+|\a_{12}|^2 ] \nn\\
&& \phantom{S_{A,k_1k_2}^\bos =}
                     - (x+|\a_{12}|)(1-x-|\a_{12}|) \log [(x+|\a_{12}|)(1-x-|\a_{12}|)] \nn\\
&& \phantom{S_{A,k_1k_2}^\bos =}
                     - (x-|\a_{12}|)(1-x+|\a_{12}|) \log [(x-|\a_{12}|)(1-x+|\a_{12}|)],
\eea
which are beyond the results in \cite{Zhang:2020dtd}.

\section{Double-magnon bound state in XXX chain}\label{appBS}

In this appendix, we discuss the R\'enyi and entanglement entropies in the double-magnon bound state, i.e.\ the class III solutions in \cite{Karbach:1998abi} and the case III solutions in this paper, in the XXX chain.
This appendix has overlaps with \cite{Molter2014Bound}.
The new things in this paper are that we give the explicit analytical expressions of the R\'enyi and entanglement entropies and we also discuss the behaviors of the results in various limits.
We show explicitly that the double-magnon bound state R\'enyi and entanglement entropies approach the single-particle R\'enyi and entanglement entropies $S_{A,k}^{(n)}$ in the strongly bound limit and approach the double-particle R\'enyi and entanglement entropies in the bosonic chain $S_{A,k^2}^{(n),\bos}$ in the loosely bound limit.

\subsection{Case IIIa solutions}

There are two subcases in the case III solutions.
For case IIIa solutions take the form $|I_1I_2\rag$ (\ref{XXXI1I2}) with (\ref{cUj1j2}), (\ref{cN}).
The actual momenta are
\be \label{XXXIIIap1p2}
p_1=\f{\pi I}{L}+\ii v, ~~
p_2=\f{\pi I}{L}-\ii v, ~~
\th=\pi+\ii L v,
\ee
with the possible values of the total Bethe number
\be
I=\td I,\td I+2,\cdots,\f{L}{2}-1,\f{3L}{2}+1,\f{3L}{2}+3,\cdots,2L-\td I.
\ee
There is the odd integer $\td I \approx \sr{L}/\pi$.
The Bethe numbers of the two magnons are
\be
I_1=\f{I-1}{2}, ~~
I_2=\f{I+1}{2}.
\ee
We have the scaled total Bethe number
\be \label{XXXIIIaiota}
\iota = \lim_{L\to+\inf} \f{I}{L}.
\ee
Using the equation (\ref{Betheequation}) in the scaling limit $L\to+\inf$, one could get
\be  \label{XXXIIIaiotalimitLtoinf}
\lim_{L\to+\inf} v = - \log \cos(\pi\io).
\ee
The parameter $v$ is in the range
\be
\f{1}{2L} \lesssim v \leq +\inf.
\ee
The parameter $1/v$ is roughly the size of the two-magnon bound state.
When $v\to0$, the two magnons are loosely bound, and when $v\to+\inf$, the two magnons are tightly bound.

Formally, we still work in the nonorthonormal basis (\ref{XXXphi0123}) with (\ref{XXXIIIap1p2}).
We get the matrices
\bea
&& \cP_{A,I_1I_2} = \f{L^2}{\cN}
\lt(\ba{cccc}
\cB &&&\\
&\f{\sinh[(L-\ell)v]}{L\sinh v}\ep^{(\ell+1)v}&-(1-x)&\\
&-(1-x)&\f{\sinh[(L-\ell)v]}{L\sinh v}\ep^{-(\ell+1)v}&\\
&&&1
\ea\rt), \nn\\
&& \cQ_{A,I_1I_2} =
\lt(\ba{cccc}
1 &&&\\
&\f{\sinh(\ell v)}{L\sinh v}\ep^{-(\ell+1)v}&x&\\
&x&\f{\sinh(\ell v)}{L\sinh v}\ep^{(\ell+1)v}&\\
&&&\cA
\ea\rt),
\eea
with
\bea \label{XXXIIIacNcAcB}
&& \cN = L \Big[ \f{\sinh[(L-1)v]}{\sinh v} - (L-1) \Big], \nn\\
&& \cA= \f{1}{L^2} \Big[ \f{\ell\sinh[(L-1)v]}{\sinh v}
                       - \f{\cosh(Lv)-\cosh[(L-2\ell)v]}{2\sinh^2v}
                       - \ell(\ell-1) \Big], \nn\\
&& \cB= \f{1}{L^2} \Big[ \f{(L-\ell)\sinh[(L-1)v]}{\sinh v}
                       - \f{\cosh(Lv)-\cosh[(L-2\ell)v]}{2\sinh^2v}
                       - (L-\ell)(L-\ell-1) \Big].
\eea
Then we obtain the matrix $\cR_{A,I_1I_2}=\cP_{A,I_1I_2}\cQ_{A,I_1I_2}$ with the eigenvalues
\bea \label{XXXIIIan1234}
&& \n_1=\f{L^2\cA}{\cN}, ~~
   \n_2=\f{L^2\cB}{\cN},\\
&& \n_{3/4} = \f{1}{\cN} \Big[ \f{\sinh(\ell v)\sinh[(L-\ell)v]}{\sinh^2v}
                             - \ell(L-\ell)
                             \pm \f{\ell\sinh[(L-\ell)v] - (L-\ell)\sinh(\ell v)}{\sinh v} \Big].\nn
\eea
The R\'enyi and entanglement entropies are
\bea
&& S_{A,I_1I_2}^{(n),\XXX} = - \f{1}{n-1} \log \Big( \sum_{i=1}^4 \n_i^n \Big), \label{SAI1I2nXXXIIIa}\\
&& S_{A,I_1I_2}^\XXX = - \sum_{i=1}^4 \n_i \log \n_i \label{SAI1I2XXXIIIa}.
\eea

In the scaling limit $L\to+\inf$, $\ell\to+\inf$ with fixed $x=\f{\ell}{L}$ and finite $v$ or infinite $v$, the size of the bound state is much smaller than the sizes of the subsystems
\be
\f1v \ll \min(\ell,L-\ell),
\ee
and the bound state behaviors like a single particle. Indeed, we have
\be \label{XXXIIIan1234x1mx00}
\n_1 = x, ~~ \n_2 = 1-x, ~~ \n_3 = 0, ~~ \n_4 = 0,
\ee
and the R\'enyi and entanglement entropies approach
\bea
&& S_{A,I_1I_2}^{(n),\XXX} = - \f{1}{n-1} \log [ x^n + (1-x)^n ] = S_{A,k}^{(n)}, \label{SAI1I2nXXXIIIasingleparticle}\\
&& S_{A,I_1I_2}^{\XXX} = -x\log x - (1-x)\log(1-x) = S_{A,k}. \label{SAI1I2XXXIIIasingleparticle}
\eea

The smallest $v$ behaves as $v=\f{u}{L}$ with fixed $u$ in the scaling limit $L\to+\inf$, the eigenvalues (\ref{XXXIIIan1234}) become
\bea
&& \n_1 = \f{2u^2x(\f{\sinh u}{u}-x)+\cosh[(1-2x)u]-\cosh u}{2u^2(\f{\sinh u}{u}-1)}, \nn\\
&& \n_2 = \f{2u^2(1-x)[\f{\sinh u}{u}-(1-x)]+\cosh[(1-2x)u]-\cosh u}{2u^2(\f{\sinh u}{u}-1)}, \nn\\
&& \n_3 = \f{[\f{\sinh (x u)}{u}+x][\f{\sinh [(1-x)u]}{u}-(1-x)]}{\f{\sinh u}{u}-1}, \nn\\
&& \n_4 = \f{[\f{\sinh (x u)}{u}-x][\f{\sinh [(1-x)u]}{u}+1-x]}{\f{\sinh u}{u}-1}.  \label{XXXIIIan1n2n3n4}
\eea
The R\'enyi and entanglement entropies are still in the forms (\ref{SAI1I2nXXXIIIa}) and (\ref{SAI1I2XXXIIIa}).
In the further limit $u\to+\inf$, we also have the eigenvalues (\ref{XXXIIIan1234x1mx00}) and the single particle R\'enyi and entanglement entropies (\ref{SAI1I2nXXXIIIasingleparticle}) and (\ref{SAI1I2XXXIIIasingleparticle}).

\begin{figure}[t]
  \centering
  \includegraphics[height=0.31\textwidth]{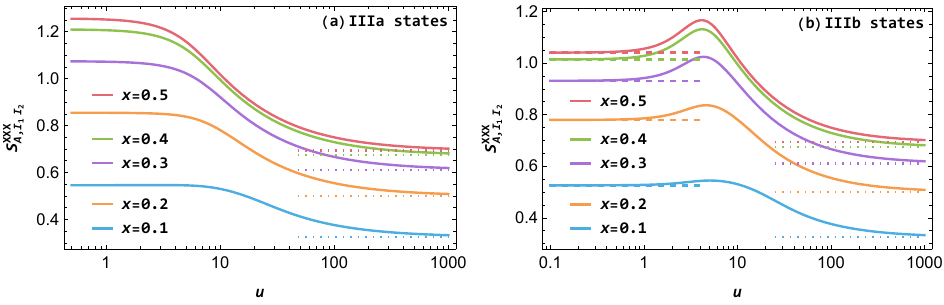}\\
  \caption{The entanglement entropy in the IIIa double-particle bound states $S_{A,I_1I_2}^\XXX$ (\ref{SAI1I2XXXIIIa}) with (\ref{XXXIIIan1n2n3n4}) (left) and the entanglement entropy in the IIIb double-particle bound states $S_{A,I_1I_2}^\XXX$ (\ref{SAI1I2XXXIIIa}) with (\ref{XXXIIIbn1n2n3n4}) (right) in the XXX chain.
  The horizonal axes $u$ is defined as $v=\f{u}{L}$.
  In the left and right panels, the dotted lines are the lower bound of the entanglement entropy $S_{A,k}=-x\log x-(1-x)\log(1-x)$.
  In the right panel, the dashed lines are $S_{A,k^2}^\bos = -x^2\log (x^2)-2x(1-x)\log[2x(1-x)]-(1-x)^2\log[(1-x)^2]$ (\ref{BosonSAk2}).}
  \label{FigureXXXboundstate}
\end{figure}

We summarize the R\'enyi and entanglement entropies in the class IIIa double-magnon bound states in the XXX chain, i.e.\ (\ref{SAI1I2nXXXIIIa}) and (\ref{SAI1I2XXXIIIa}) with (\ref{XXXIIIan1234}) and (\ref{XXXIIIacNcAcB}), in table~\ref{tableIIIa}.
We show entanglement entropy in the IIIa double-particle bound states $S_{A,I_1I_2}^\XXX$ (\ref{SAI1I2XXXIIIa}) with (\ref{XXXIIIan1n2n3n4}) in the left panel of figure~\ref{FigureXXXboundstate}.

\begin{table}[t]
  \centering
  \begin{tabular}{|c|c|c|c|c|c|}
  \hline
  \mr{2}{IIIa states}       & \mc{2}{c|}{$v=\f{u}{L}$}  & \mr{2}{$\cdots$} & \mr{2}{$v$ finite} & \mr{2}{$v\to+\inf$}  \\ \cline{2-3}
                            & $u$ finite & $u\to+\inf$  &                  &            &              \\ \hline
   \mr{2}{$S_{A,I_1I_2}^{(n),\XXX}$} & \mc{2}{c|}{(\ref{SAI1I2nXXXIIIa}) and (\ref{SAI1I2XXXIIIa}) with (\ref{XXXIIIan1n2n3n4})} & \mc{3}{c|}{} \\ \cline{3-3}
   & \mc{1}{c|}{} & \mc{4}{c|}{$S_{A,k}^{(n)}$}\\\hline
  \end{tabular}
  \caption{The R\'enyi and entanglement entropies in case IIIa double-magnon bound states of XXX chain.}
  \label{tableIIIa}
\end{table}

\subsection{Case IIIb solutions}

For case IIIb solutions there are
\be \label{XXXIIIbp1p2}
p_1=\f{\pi I}{L}+\ii v, ~~
p_2=\f{\pi I}{L}-\ii v, ~~
\th=\ii L v,
\ee
with the possible values of the total Bethe number
\be
I=2,4,\cdots,\f{L}{2},\f{3L}{2},\cdots,2L-2,\ee
and the Bethe numbers of the two quasiparticles
\be
I_1=I_2=\f{I}{2}.
\ee
We still have the scaled total Bethe number defined as (\ref{XXXIIIaiota}) and behave in the scaling limit as (\ref{XXXIIIaiotalimitLtoinf}).
The parameter $v$ is in the range
\be
\f{2\pi^2}{L^2} \leq v \leq +\inf.
\ee
The parameter $1/v$ is still the size of the two-magnon bound state.

In the nonorthonormal basis (\ref{XXXphi0123}) with (\ref{XXXIIIbp1p2}), we get the matrices
\bea
&& \cP_{A,I_1I_2} = \f{L^2}{\cN}
\lt(\ba{cccc}
\cB &&&\\
&\f{\sinh[(L-\ell)v]}{L\sinh v}\ep^{(\ell+1)v}&1-x&\\
&1-x&\f{\sinh[(L-\ell)v]}{L\sinh v}\ep^{-(\ell+1)v}&\\
&&&1
\ea\rt), \nn\\
&& \cQ_{A,I_1I_2} =
\lt(\ba{cccc}
1 &&&\\
&\f{\sinh(\ell v)}{L\sinh v}\ep^{-(\ell+1)v}&x&\\
&x&\f{\sinh(\ell v)}{L\sinh v}\ep^{(\ell+1)v}&\\
&&&\cA
\ea\rt),
\eea
with
\bea \label{XXXIIIbcNcAcB}
&& \cN = L \Big[ \f{\sinh[(L-1)v]}{\sinh v} + L-1 \Big], \nn\\
&& \cA= \f{1}{L^2} \Big[ \f{\ell\sinh[(L-1)v]}{\sinh v}
                       - \f{\cosh(Lv)-\cosh[(L-2\ell)v]}{2\sinh^2v}
                       + \ell(\ell-1) \Big], \nn\\
&& \cB= \f{1}{L^2} \Big[ \f{(L-\ell)\sinh[(L-1)v]}{\sinh v}
                       - \f{\cosh(Lv)-\cosh[(L-2\ell)v]}{2\sinh^2v}
                       + (L-\ell)(L-\ell-1) \Big].
\eea
Then we obtain the matrix $\cR_{A,I_1I_2}=\cP_{A,I_1I_2}\cQ_{A,I_1I_2}$ with the eigenvalues
\bea \label{XXXIIIbn1234}
&& \n_1=\f{L^2\cA}{\cN}, ~~
   \n_2=\f{L^2\cB}{\cN}, \nn\\
&& \n_{3/4} = \f{1}{\cN} \Big[ \f{\sinh(\ell v)\sinh[(L-\ell)v]}{\sinh^2v}
                             + \ell(L-\ell)
                             \pm \f{\ell\sinh[(L-\ell)v] + (L-\ell)\sinh(\ell v)}{\sinh v} \Big].
\eea
The R\'enyi and entanglement entropies are in the forms (\ref{SAI1I2nXXXIIIa}) and (\ref{SAI1I2XXXIIIa}).

In the scaling limit $L\to+\inf$, $\ell\to+\inf$ with fixed $x=\f{\ell}{L}$ and finite $v$ or infinite $v$, we have (\ref{XXXIIIan1234x1mx00}) and the R\'enyi and entanglement entropies approach (\ref{SAI1I2nXXXIIIasingleparticle}) and (\ref{SAI1I2XXXIIIasingleparticle}).

The smallest $v$ behaves as $v=\f{w}{L^2}$ with fixed $w\geq2\pi^2$ in the scaling limit $L\to+\inf$, and the eigenvalues (\ref{XXXIIIbn1234}) become
\be \label{XXXIIIbn1234x21mx22x1mx}
\n_1=x^2, ~~
\n_2=(1-x)^2, ~~
\n_3=2x(1-x),~~
\n_4=0.
\ee
The R\'enyi and entanglement entropies are
\bea
&& S_{A,I_1I_2}^{(n),\XXX} = - \f{1}{n-1}\log\{x^{2n}+[2x(1-x)]^n+(1-x)^{2n}\} = S_{A,k^2}^{(n),\bos}, \label{SAI1I2nXXXIIIbdoubleparticle}\\
&& S_{A,I_1I_2}^{\XXX} = -x^2\log (x^2)-2x(1-x)\log[2x(1-x)]-(1-x)^2\log[(1-x)^2] = S_{A,k^2}^\bos, \label{SAI1I2XXXIIIbdoubleparticle}
\eea

If $v$ behaves as $v=\f{u}{L}$ with fixed $u\geq2\pi^2$ in the scaling limit $L\to+\inf$, the eigenvalues (\ref{XXXIIIbn1234}) become
\bea \label{XXXIIIbn1n2n3n4}
&& \n_1 = \f{2u^2x(\f{\sinh u}{u}+x)+\cosh[(1-2x)u]-\cosh u}{2u^2(\f{\sinh u}{u}+1)}, \nn\\
&& \n_2 = \f{2u^2(1-x)[\f{\sinh u}{u}+1-x]+\cosh[(1-2x)u]-\cosh u}{2u^2(\f{\sinh u}{u}+1)}, \nn\\
&& \n_3 = \f{[\f{\sinh (x u)}{u}+x][\f{\sinh [(1-x)u]}{u}+1-x]}{\f{\sinh u}{u}+1}, \nn\\
&& \n_4 = \f{[\f{\sinh (x u)}{u}-x][\f{\sinh [(1-x)u]}{u}-(1-x)]}{\f{\sinh u}{u}+1}.
\eea
The R\'enyi and entanglement entropies are in the forms (\ref{SAI1I2nXXXIIIa}) and (\ref{SAI1I2XXXIIIa}).
In the further limit $u\to+\inf$, we have the eigenvalues (\ref{XXXIIIan1234x1mx00}) and the single particle R\'enyi and entanglement entropies (\ref{SAI1I2nXXXIIIasingleparticle}) and (\ref{SAI1I2XXXIIIasingleparticle}).
In the further limit $u\to0$, we have the eigenvalues (\ref{XXXIIIbn1234x21mx22x1mx}) and the R\'enyi and entanglement entropies (\ref{SAI1I2nXXXIIIbdoubleparticle}) and (\ref{SAI1I2XXXIIIbdoubleparticle}).

We summarize the R\'enyi and entanglement entropies in the class IIIb double-magnon bound states in the XXX chain, i.e.\ (\ref{SAI1I2nXXXIIIa}) and (\ref{SAI1I2XXXIIIa}) with (\ref{XXXIIIbn1234}) and (\ref{XXXIIIbcNcAcB}), in table~\ref{tableIIIb}.
We show entanglement entropy in the IIIb double-particle bound states $S_{A,I_1I_2}^\XXX$ (\ref{SAI1I2XXXIIIa}) with (\ref{XXXIIIbn1n2n3n4}) in the right panel of figure~\ref{FigureXXXboundstate}.
It is interesting to note that with fixed $x$ the entanglement entropy does not decrease monotonically with the increase of $u$.

\begin{table}[t]
  \centering
  \begin{tabular}{|c|c|c|c|c|c|c|c|c|}
  \hline
  \mr{2}{IIIb states}               & \mr{2}{$v=\f{w}{L^2}$} & \mr{2}{$\cdots$} & \mc{3}{c|}{$v=\f{u}{L}$}            & \mr{2}{$\cdots$} & \mr{2}{$v$ finite} & \mr{2}{$v\to+\inf$}  \\ \cline{4-6}
                                    &                        &                  & $u\to0$ & $u$ finite & $u\to+\inf$  &                  &            &              \\ \hline
  \mr{2}{$S_{A,I_1I_2}^{(n),\XXX}$} & \mc{2}{c|}{}                              & \mc{3}{c|}{(\ref{SAI1I2nXXXIIIa}) and (\ref{SAI1I2XXXIIIa}) with (\ref{XXXIIIbn1n2n3n4})}                        & \mc{3}{c|}{} \\ \cline{4-4} \cline{6-6}
                                    & \mc{3}{c|}{$S_{A,k^2}^{(n),\bos}$}                  &            & \mc{4}{c|}{$S_{A,k}^{(n)}$}\\\hline
\end{tabular}
  \caption{The R\'enyi and entanglement entropies in case IIIb double-magnon bound states of XXX chain.}
  \label{tableIIIb}
\end{table}


\providecommand{\href}[2]{#2}\begingroup\raggedright\endgroup

\end{document}